\documentclass[11pt,a4paper]{amsart}
\usepackage[centertags]{amsmath}
\usepackage{amsfonts}
\usepackage{amssymb}
\usepackage{amsthm}
\usepackage{newlfont}
\usepackage{a4}
\usepackage{enumerate}

\theoremstyle{definition}
\newtheorem{defn}{Definition}[section]
\newtheorem{thm}[defn]{Theorem}
\newtheorem{lemma}[defn]{Lemma}

\newtheorem{cor}[defn]{Corollary}
\theoremstyle{remark}
\newtheorem{example}{Example}[section]

\usepackage[centertags]{amsmath}
\usepackage{amsfonts}
\usepackage{amssymb}
\usepackage{amsthm}
\usepackage{bbm}
\usepackage{newlfont}
\usepackage{a4}
\usepackage{enumerate}

\newlength{\defbaselineskip}
\newcommand{\setlinespacing}[1]%
           {\setlength{\baselineskip}{#1 \defbaselineskip}}

\newcommand{\map}{\rightarrow}
\newcommand{\der}{\operatorname{der}}
\newcommand{\coc}{\operatorname{coc}}
\newcommand{\ad}{\operatorname{ad}}

\newcommand{\slp}{\mathop{\mathrm {sl} }\nolimits}

\newcommand{\End}{\operatorname{End}}

\newcommand{\g}{\operatorname{g}}

\newcommand{\fa}{\operatorname{\psi}}
\newcommand{\fb}{\operatorname{\phi}}
\newcommand{\fc}{\operatorname{\phi^0}}

\newcommand{\el}{{\cal L}}

\newcommand{\q}{\quad}

\renewcommand{\epsilon}{\varepsilon}

\newcommand{\ep}{\varepsilon}
\newcommand{\la}{\lambda}
\newcommand{\al}{\alpha}
\renewcommand{\rho}{\varrho}
\renewcommand{\phi}{\varphi}

\newcommand{\N}{{\mathbb N}}

\newcommand{\Com}{{\mathbb C}}

\newcommand{\Z}{\mathbb{Z}}

\newcommand{\wt}{\widetilde}
\newcommand{\gc}[2]{\coc_{( #1 )}\, #2}
\newcommand{\gd}[2]{\der_{( #1 )}\, #2}

\begin{document}

\title[Twisted Cocycles of Lie algebras]{Twisted cocycles of Lie Algebras
and Corresponding Invariant Functions}

\author{Ji\v{r}\'\i\ Hrivn\'ak, Petr Novotn\'y}%

\date{\today}

\begin{center}\Large\bf
Twisted Cocycles of Lie Algebras
and Corresponding Invariant Functions
\end{center}
\bigskip
\begin{center}\large
Ji\v{r}\'\i\ Hrivn\'ak\footnote{Corresponding author: Tel.: +420 2 24358351; fax: +420 2 22320861 \newline {\it E-mail address:} jiri.hrivnak@fjfi.cvut.cz (J. Hrivn\'ak)}, Petr Novotn\'y
\end{center}
\bigskip
\begin{center}
Department of Physics,
Faculty of Nuclear sciences and Physical Engineering, Czech
Technical University, B\v{r}ehov\'a 7, 115 19 Prague 1, Czech
Republic
\end{center}
\vspace{24pt}
\hrule\vspace{8pt}
{\footnotesize
ABSTRACT. We consider finite-dimensional complex Lie algebras. Using certain
complex parameters we generalize the concept of cohomology
cocycles of Lie algebras. A special case is generalization of
1--cocycles with respect to the adjoint representation -- so called
$(\alpha,\beta,\gamma)$--derivations. Parametric sets of spaces of
cocycles allow us to define complex functions which are invariant
under Lie isomorphisms. Such complex functions thus represent
useful invariants -- we show how they classify three and
four-dimensional Lie algebras as well as how they apply to some
eight-dimensional one-parametric nilpotent continua of Lie
algebras. These functions also provide necessary criteria for
existence of 1--parametric continuous contraction.
}\vspace{8pt}
\hrule
\vspace{24pt}

\section{Introduction}
The search for a new concept of invariant characteristics of Lie
algebras led to the definition of
$(\alpha,\beta,\gamma)$--derivations in~\cite{NHd}. Consider a
complex Lie algebra $\el$ and its derivation $D\in \der \el$, i.e.
a linear operator $D\in\End\el$ satisfying for all $x,y\in \el$
the equation
 $D[x,y]=[Dx,y]+[x,Dy]$.
For fixed $\alpha,\beta,\gamma\in \Com$ is a linear operator $A\in
\End\el$ called an {\it $(\alpha,\beta,\gamma)$--derivation} if
for all $x,y\in \el$ the equation
\begin{equation}\label{gender}
\alpha A[x,y]=\beta [Ax,y]+\gamma [x, Ay]
\end{equation}
holds~\cite{NHd}. In this article we denote the set of all
operators satisfying~(\ref{gender}) by
$\gd{\alpha,\beta,\gamma}{\el}$. Investigating the spaces
$\gd{\alpha,\beta,\gamma}{\el}$, various Jordan and Lie operator
algebras were obtained. In~\cite{NHd} was also shown that the
dimensions of these operator algebras and, in fact, the dimensions of
all spaces $\gd{\alpha,\beta,\gamma}{\el}$
as well as their intersections
form invariant
characteristics of Lie algebras. The invariance of these
dimensions enabled us to define so called {\it invariant function
$\fa\el$} by the relation
\begin{equation}\label{fa}
  \fa\el(\alpha)\equiv \dim \gd{\alpha,1,1}{\el}.
\end{equation}
This complex function $\fa\el$ turned out to be very valuable: it
was used to classify all complex three--dimensional indecomposable
Lie algebras. More significantly, it resolved some parametric
continua of nilpotent Lie algebras.

Prior to invention of the invariant function $\fa\el$, the
isomorphism problem for algebras inside nilpotent parametric
continuum had to be solved explicitly. If algebras inside a
nilpotent parametric continuum are considered then all well known
characteristics such as dimensions of {\it derived}, {\it lower
central} and {\it upper central sequences}, Lie {\it algebra of
derivations}, {\it radical}, {\it nilradical}, {\it ideals}, {\it
subalgebras} \cite{Jacobson,ide} and {\it megaideals} \cite{Pop1}
naturally coincide. Moreover, due to Engel's theorem none of the
'trace' invariants based on the adjoint representation such
as $C_{pq}$~\cite{Bur1} or $\chi_i$~\cite{AY} exists in this case. The nilpotent parametric continua appeared
for example in~\cite{HN2} where all graded contractions~\cite{PdeM,MP} of the Pauli graded $\slp(3,\Com)$~\cite{PZ1} were found. The invariant function $\fa\el$ was successfully applied in~\cite{NHd} to some of the nilpotent parametric continua from~\cite{HN2}. This invariant function {\it did not}, however, resolve {\it all} nilpotent continua which were discussed in~\cite{HN2}.

Often, the search of new invariant characteristics of Lie algebras is motivated by the classification of degenerations~\cite{Bur1,Bur2,Bur3} or 1-parametric continuous contractions~\cite{Nes}. The applicability of the invariant function $\fa\el$ (or its possible generalizations) in this field remained an open problem.

Precisely these facts motivated the work undertaken in this article. The idea of finding new invariant characteristics of Lie algebras via generalizing standard Chevalley cohomology cocycles is followed. New invariant functions are defined and applied not only to some cases from~\cite{HN2} but to all four-dimensional complex Lie algebras. Application of these generalized cocycles to 1-parametric continuous contractions is discussed. It is shown that these new invariants of Lie algebras constitute an essential tool for resolving parametric continua of nilpotent Lie algebras and provide powerful necessary contraction criteria.

In Section 2, twisted cocycles of Lie algebras are introduced and two--dimensional twisted cocycles of the adjoint representation are investigated in detail. It is shown that there are sixteen cases of these two-dimensional cocycles which can be described by four complex parameters. 

In Section 3, the invariant functions
$\fb$ and $\fc$ are defined and their behavior on
low--dimensional Lie algebras demonstrated. The invariant
functions $\fa$ and $\fb$ are used to classify all three and
four--dimensional complex Lie algebras. New algorithm for the
identification of a four--dimensional complex Lie algebra is also
formulated.

In Section 4, possible application of the invariant
functions $\fa,\fb,\fc$ to contractions is considered. Necessary
criterion for existence of a 1-parametric continuous contraction is formulated.
The invariant function $\fa$ is used to classify continuous
contractions among three--dimensional Lie algebras. Application of the invariant functions on
nilpotent parametric continua of Lie algebras resulting from
contractions of the Pauli graded $\slp(3,\Com)$ is demonstrated.

In Appendix A, the tables of the invariant functions
$\fa,\,\fb,\,\fc$ for two, three and four--dimensional Lie
algebras are listed.

In Appendix B, proof of the classification Theorem \ref{class4dim} is located.
\section{Twisted Cocycles of Lie Algebras}\label{CHtwi}
Let $\el$ be an arbitrary complex Lie algebra and $(V,f)$ its
representation. We denote by $C^q(\el,V)$ the vector space of all
$V$--cochains of dimension $q$ for $q\in \N$ and $C^0(\el,V)=V$. We
generalize the notions of cocycles analogously to
$(\alpha,\beta,\gamma)$--derivations. Let $\kappa=(\kappa_{ij})$ be a $(q+1)\times(q+1)$ complex symmetric
matrix. We call $c\in C^q(\el,V)$, $q\in\N$ for which
\begin{align}\label{koho1t}
\nonumber 0 &= \sum_{i=1}^{q+1}
(-1)^{i+1}\kappa_{ii}f(x_i)c(x_1,\dots,\hat{x_i},\dots,x_{q+1})+\\
 &+ \sum_{\begin{smallmatrix}
  i,j=1 \\
  i<j
\end{smallmatrix}}^{q+1} (-1)^{i+j}\kappa_{ij} c([x_i,x_j],x_1,\dots,\hat{x_i},\dots,\hat{x_j},\dots,x_{q+1})
\end{align} a {\bf
$\kappa-$twisted cocycle} or shortly $\kappa-$cocycle of dimension
$q$ corresponding to $(V,f)$. The symbol $\hat{x_i}$ means that the term $x_i$ is omitted. The set of all $\kappa-$cocycles of
dimension $q$ is denoted by $Z^{q} (\el,f,\kappa)$ and it is clear that
$Z^{q} (\el,f,\kappa)$ is a linear subspace of $C^q(\el,V)$. We
observe that for any $\varepsilon \in \Com \backslash\{0\}$ and
$q\in \N$ it holds:
\begin{equation}\label{vlakoho1}
Z^{q} (\el,f,\kappa)=Z^{q} (\el,f,\varepsilon\kappa).
\end{equation}

\subsection{Two--dimensional Twisted Cocycles of the Adjoint
Representation}
The definition of
$(\alpha,\beta,\gamma)$--derivations is now included in the
definition of twisted cocycles. Considering the adjoint
representation and its one-dimensional twisted cocycles, we immediately have:
\begin{equation}\label{SPEC}
Z^{1} \left(\el,\ad_\el,\left(\begin{smallmatrix}
  \beta &  \alpha \\
   \alpha & \gamma
\end{smallmatrix}\right) \right)=\gd{\alpha,\beta,\gamma}\el.
\end{equation}
In this section we investigate in detail the space $Z^{2} (\el,\ad_\el,\kappa)$. For this purpose it may be more convenient to use different notation, analogous to that of derivations, defined by
\begin{equation}\label{kohonot}
\gc{\alpha_1,\alpha_2,\alpha_3,\beta_1,\beta_2,\beta_3}\el=Z^{2}
\left(\el,\ad_\el,\left(\begin{smallmatrix}
  \beta_1 &   \alpha_2 &  \alpha_3 \\
   \alpha_2 & \beta_3 &  \alpha_1 \\
   \alpha_3 &  \alpha_1 & \beta_2
\end{smallmatrix}\right) \right),
\end{equation}
i.~e. in the space
$\gc{\alpha_1,\alpha_2,\alpha_3,\beta_1,\beta_2,\beta_3}\el$ are
such $B \in C^2(\el,\el)$ which for all $x,y,z \in\el$ satisfy
\begin{align}\label{defb}
0&=\alpha_1 B(x, [y,z])+\alpha_2 B(z, [x,y])+\alpha_3 B(y, [z,x])
\nonumber\\ &+ \beta_1 [x, B ( y,z)]+ \beta_2 [z, B (
x,y)]+\beta_3 [y, B ( z,x)].
\end{align}
Six permutations of the variables $x,y,z \in\el$ in the defining
equation (\ref{defb}) give
\begin{lemma}
 Let $\el$ be a complex Lie algebra. Then for any $\alpha_1, \alpha_2,\alpha_3,\beta_1,\beta_2,\beta_3 \in \Com$
are all the following six spaces equal:
\begin{enumerate}
\item $\gc{\alpha_1,\alpha_2,\alpha_3,\beta_1,\beta_2,\beta_3}\el$
\item $\gc{\alpha_3,\alpha_1,\alpha_2,\beta_3,\beta_1,\beta_2}\el$
\item $\gc{\alpha_2,\alpha_3,\alpha_1,\beta_2,\beta_3,\beta_1}\el$
\item $\gc{\alpha_1,\alpha_3,\alpha_2,\beta_1,\beta_3,\beta_2}\el$
\item $\gc{\alpha_2,\alpha_1,\alpha_3,\beta_2,\beta_1,\beta_3}\el$
\item $\gc{\alpha_3,\alpha_2,\alpha_1,\beta_3,\beta_2,\beta_1}\el$
\end{enumerate}
\end{lemma}

\begin{lemma}\label{vlakoho} Let $\el$ be a complex Lie algebra. Then for any $\alpha_1, \alpha_2,\alpha_3,\beta_1,\beta_2,\beta_3 \in \Com$
is the space $
\gc{\alpha_1,\alpha_2,\alpha_3,\beta_1,\beta_2,\beta_3}\el$ equal
to all of the following:
\begin{enumerate}
\item$\gc{\alpha_1+\alpha_3,\alpha_2+\alpha_1,\alpha_3+\alpha_2,\beta_1+\beta_3,\beta_2+\beta_1,\beta_3+\beta_2}\el\,
\cap\,\gc{\alpha_1-\alpha_3,\alpha_2-\alpha_1,\alpha_3-\alpha_2,\beta_1-\beta_3,\beta_2-\beta_1,\beta_3-\beta_2}\el$
\item$\gc{0,\alpha_2-\alpha_3,\alpha_3-\alpha_2,0,\beta_2-\beta_3,\beta_3-\beta_2}\el\,
\cap\,\gc{2\alpha_1,\alpha_2+\alpha_3,\alpha_3+\alpha_2,2\beta_1,\beta_2+\beta_3,\beta_3+\beta_2}\el$
\item$\gc{0,\alpha_1-\alpha_2,\alpha_2-\alpha_1,0,\beta_1-\beta_2,\beta_2-\beta_1}\el\,
\cap\,\gc{2\alpha_3,\alpha_1+\alpha_2,\alpha_2+\alpha_1,2\beta_3,\beta_1+\beta_2,\beta_2+\beta_1}\el$
\item$\gc{0,\alpha_3-\alpha_1,\alpha_1-\alpha_3,0,\beta_3-\beta_1,\beta_1-\beta_3}\el\,
\cap\,\gc{2\alpha_2,\alpha_3+\alpha_1,\alpha_1+\alpha_3,2\beta_2,\beta_3+\beta_1,\beta_1+\beta_3}\el$

\end{enumerate}
\end{lemma}

\begin{proof} Suppose $\alpha_1,\alpha_2,\alpha_3,\beta_1,\beta_2,\beta_3 \in \Com$ are given. We demonstrate the proof on the case
1. -- the proof of the other cases is analogous. Let $B \in
\gc{\alpha_1,\alpha_2,\alpha_3,\beta_1,\beta_2,\beta_3}\el$; then
for arbitrary $x,y,z\in \el$ we have
    \begin{eqnarray}
    0&=\alpha_1 B(x, [y,z])+\alpha_2 B(z, [x,y])+\alpha_3 B(y, [z,x])
\nonumber\\ &+ \beta_1 [x, B ( y,z)]+ \beta_2 [z, B (
x,y)]+\beta_3 [y, B ( z,x)]\label{trik3} \\
    0&=\alpha_1 B(z, [x,y])+\alpha_2 B(y, [z,x])+\alpha_3 B(x, [y,z])
\nonumber\\ &+ \beta_1 [z, B ( x,y)]+ \beta_2 [y, B (
z,x)]+\beta_3 [x, B ( y,z)]. \label{trik4}
    \end{eqnarray}
    By adding and subtracting equations (\ref{trik3}) and (\ref{trik4}) we obtain
     \begin{eqnarray}
    0&=(\alpha_1+\alpha_3) B(x, [y,z])+(\alpha_2+\alpha_1) B(z, [x,y])+(\alpha_3 +\alpha_2 )B(y, [z,x])
\nonumber\\ &+ (\beta_1+\beta_3) [x, B ( y,z)]+ (\beta_2+\beta_1)
[z, B ( x,y)]+(\beta_3+\beta_2) [y, B ( z,x)]\label{trik5} \\
    0&=(\alpha_1-\alpha_3) B(x, [y,z])+(\alpha_2-\alpha_1) B(z, [x,y])+(\alpha_3 -\alpha_2 )B(y, [z,x])
\nonumber\\ &+ (\beta_1-\beta_3) [x, B ( y,z)]+ (\beta_2-\beta_1)
[z, B ( x,y)]+(\beta_3-\beta_2) [y, B ( z,x)]\label{trik6}
    \end{eqnarray}
    and thus $
\gc{\alpha_1,\alpha_2,\alpha_3,\beta_1,\beta_2,\beta_3}\el$ is the
subset of the intersection 1. Similarly, starting with equations
(\ref{trik5}), (\ref{trik6}) we obtain equations (\ref{trik3}),
(\ref{trik4}) and the remaining inclusion is proven.
\end{proof}
We show in the following theorem that four parameters are sufficient
for the description of all spaces of two--dimensional twisted
cocycles.
\begin{thm}\label{klass2} Let $\el$ be a Lie algebra. Then for any $\alpha_1,\alpha_2,\alpha_3,\beta_1,\beta_2,\beta_3
 \in\Com$ there exist $\alpha,\beta,\gamma,\delta\in \Com$ such that the subspace
$\gc{\alpha_1,\alpha_2,\alpha_3,\beta_1,\beta_2,\beta_3}\el
\subset C^2(\el,\el)$ is equal to some of the following sixteen
subspaces:
\begin{enumerate}
 \item $\gc{\alpha,0,0,\beta,0,0}{\el}$;
  $\gc{\alpha,0,0,\beta,1,-1}{\el}$;
  $\gc{\alpha,1,-1,\beta,0,0}{\el}$;
  $\gc{\alpha,\beta,-\beta,\gamma,1,-1}{\el}$
 \item $\gc{\alpha,0,0,\beta,1,0}{\el}$;
  $\gc{\alpha,0,0,\beta,1,1}{\el}$;
  $\gc{\alpha,\beta,-\beta,\gamma,1,0}{\el}$;
  $\gc{\alpha,1,-1,\beta,1,1}{\el}$
 \item $\gc{\alpha,1,0,\beta,0,0}{\el}$;
  $\gc{\alpha,1,1,\beta,0,0}{\el}$;
  $\gc{\alpha,1,0,\beta,\gamma,-\gamma}{\el}$;
  $\gc{\alpha,1,1,\beta,1,-1}{\el}$
 \item $\gc{\alpha,\beta,\gamma,\delta,1,0}{\el}$;
  $\gc{\alpha,\beta+1,\beta-1,\gamma,1,1}{\el}$;
  $\gc{\alpha,1,1,\beta,\gamma+1,\gamma-1}{\el}$;
  $\gc{\alpha,\beta,\beta,\gamma,1,1}{\el}$
\end{enumerate}
\end{thm}
\begin{proof}

 Suppose $\alpha_2 + \alpha_3 = 0 $ and $\beta_2 + \beta_3 = 0 $. Then the following four
    cases are possible:
    \begin{enumerate}
    \item $\alpha_2=-\alpha_3=0$ and $\beta_2=-\beta_3=0$. In this
    case we have
    $$\gc{\alpha_1,\alpha_2,\alpha_3,\beta_1,\beta_2,\beta_3}\el  =
    \gc{\alpha_1,0,0,\beta_1,0,0}\el.$$
    \item $\alpha_2=-\alpha_3=0$ and $\beta_2=-\beta_3\neq 0$. In this
    case we have:
\begin{align*}
\gc{\alpha_1,\alpha_2,\alpha_3,\beta_1,\beta_2,\beta_3}\el=&\gc{\alpha_1,0,0,\beta_1,\beta_2,\beta_3}\el\\
=&
\gc{0,0,0,0,\beta_2-\beta_3,\beta_3-\beta_2}\el\cap\gc{2\alpha_1,0,0,2\beta_1,0,0}\el\\
=&\gc{0,0,0,0,2,-2}\el\cap\gc{2\alpha_1,0,0,2\beta_1,0,0}\el\\
=&\gc{\alpha_1,0,0,\beta_1,1,-1}\el
\end{align*}
\item $\alpha_2=-\alpha_3\neq0$ and $\beta_2=-\beta_3= 0$. In this
    case we have:
\begin{align*}
\gc{\alpha_1,\alpha_2,\alpha_3,\beta_1,\beta_2,\beta_3}\el=&\gc{\alpha_1,\alpha_2,\alpha_3,\beta_1,0,0}\el\\
=&
\gc{0,\alpha_2-\alpha_3,\alpha_3-\alpha_2,0,0,0}\el\cap\gc{2\alpha_1,0,0,2\beta_1,0,0}\el\\
=&\gc{0,2,-2,0,0,0}\el\cap\gc{2\alpha_1,0,0,2\beta_1,0,0}\el\\
=&\gc{\alpha_1,1,-1,\beta_1,0,0}\el
\end{align*}
\item $\alpha_2=-\alpha_3\neq0$ and $\beta_2=-\beta_3\neq 0$. In this
    case we have:
\begin{align*}
\gc{\alpha_1,\alpha_2,\alpha_3,\beta_1,\beta_2,\beta_3}\el=&
\gc{0,\alpha_2-\alpha_3,\alpha_3-\alpha_2,0,\beta_2-\beta_3,\beta_3-\beta_2}\el\cap\gc{2\alpha_1,0,0,2\beta_1,0,0}\el\\
=&\gc{0,2\frac{\alpha_2-\alpha_3}{\beta_2-\beta_3},-2\frac{\alpha_2-\alpha_3}{\beta_2-\beta_3},0,2,-2}\el\cap\gc{2\alpha_1,0,0,2\beta_1,0,0}\el\\
=&\gc{\alpha_1,\frac{\alpha_2-\alpha_3}{\beta_2-\beta_3},-\frac{\alpha_2-\alpha_3}{\beta_2-\beta_3},\beta_1,1,-1}\el
\end{align*}
 \end{enumerate}
Discussion of the three remaining cases of the values of $\alpha_2 + \alpha_3$ and $\beta_2 + \beta_3$ is similar to the previous case and we omit it.

\end{proof}
It may be more convenient, sometimes, to use different
distribution of the cocycle spaces
than in Theorem~\ref{klass2}. Henceforth, we investigate mainly the cocycle
space $\gc{1,1,1,\la,\la,\la}\el$ which for $\la\neq 0$ fits in
the class $ \gc{\alpha,\beta,\beta,\gamma,1,1}\el$, with
$\alpha=\beta=1/\la,\,\gamma=1$. For $\la=0$, the space
$\gc{1,1,1,0,0,0}\el$ is a special case of the space
$\gc{\al,1,1,\beta,0,0}\el$, with $\al=1,\beta=0$. We also put
$\al=0,\,\beta=1,\,\gamma=\la$ into
$\gc{\alpha,\beta,\beta,\gamma,1,1}\el$ and investigate the space
$\gc{0,1,1,\la,1,1}\el$.

\section{Invariant Functions}

\begin{thm}\label{tvr1koho}
Let $g:\el \rightarrow \widetilde{\el}$ be an isomorphism of Lie
algebras $\el$ and $\widetilde{\el}$. Then the mapping $\rho
:C^q(\el,\el)\rightarrow C^q(\wt\el,\wt\el),\,q\in \N$ defined for
all $c\in C^q(\el,\el)$ and all $x_1,\dots,x_q\in \wt\el$ by $$
(\rho c)(x_1,\dots,x_q) = g c(g^{-1}x_1,\dots,g^{-1}x_q) $$ is an
isomorphism of vector spaces $C^q(\el,\el)$ and $
C^q(\wt\el,\wt\el)$. For any complex symmetric $(q+1)$--square
matrix $\kappa$
 $$\rho(Z^{q} (\el,\ad_\el,\kappa)) =
Z^{q} (\wt\el,\ad_{\wt\el},\kappa)$$ holds.
\end{thm}
\begin{proof}
Suppose we have $g:\el \rightarrow \wt\el$ such that for all
$x,y\in \widetilde{\el}$ $$[x,y]_{ \widetilde{\el}}=g
[g^{-1}x,g^{-1}y ]_{\el} $$ holds. It is clear that the map $\rho
:C^q(\el,\el)\rightarrow C^q(\wt\el,\wt\el),\,q\in \N$ is linear
and bijective, i.~e. it is an isomorphism of these vector spaces.
By putting $f=\ad_\el$ and rewriting definition (\ref{koho1t})  we
have for $c\in Z^{q} (\el,\ad_\el,\kappa)$ and all
$x_1,\dots,x_q\in \wt\el$
\begin{align}
\nonumber 0 &= \sum_{i=1}^{q+1}
(-1)^{i+1}\kappa_{ii}[g^{-1}x_i,c(g^{-1}x_1,\dots,\widehat{g^{-1}x_i},\dots,g^{-1}x_{q+1})]_{\el}+\\
\nonumber &+ \sum_{\begin{smallmatrix}
  i,j=1 \\
  i<j
\end{smallmatrix}}^{q+1} (-1)^{i+j}\kappa_{ij} c([g^{-1}x_i,g^{-1}x_j]_\el,g^{-1}x_1,\dots,\widehat{g^{-1}x_i},\dots,\widehat{g^{-1}x_j},\dots,g^{-1}x_{q+1})
\end{align}
Applying the mapping $g$ on this equation and taking into account
that $\kappa_{ij}\in \Com$ one has
\begin{align}\label{koho1adwt}
\nonumber 0 &= \sum_{i=1}^{q+1} (-1)^{i+1}\kappa_{ii}[x_i,(\rho
c)(x_1,\dots,\hat{x_i},\dots,x_{q+1})]_{\wt\el}+\\ \nonumber &+
\sum_{\begin{smallmatrix}
  i,j=1 \\
  i<j
\end{smallmatrix}}^{q+1} (-1)^{i+j}\kappa_{ij} (\rho c)([x_i,x_j]_{\wt\el},x_1,\dots,\hat{x_i},\dots,\hat{x_j},\dots,x_{q+1})
\end{align}
i.~e. $\rho c\in Z^{q} (\wt\el,\ad_{\wt\el},\kappa)$.
\end{proof}
\begin{cor}\label{invdim}
For any $q\in \N$ and any complex symmetric $(q+1)$--square matrix
$\kappa$ is the dimension of the vector space
$Z^{q}(\el,\ad_\el,\kappa)$ an invariant characteristic of Lie
algebras.
\end{cor}
Sixteen parametric spaces in Theorem~\ref{klass2} allow us to
define sixteen invariant functions of up to four variables.
However, a complete analysis of possible outcome is beyond the
scope of this work. Rather empirically, following calculations in
dimension four and eight, we pick up two one--parametric sets of
vector spaces to define two new invariant functions of a
$n$--dimensional Lie algebra $\el$. We call functions
$\fb\el,\fc\el:\Com \rightarrow \{0,1,\dots,n^2(n-1)/2\}$ defined
by the formulas
\begin{align}
(\fb\el)(\alpha) =& \dim\gc{1,1,1,\alpha,\alpha,\alpha}{\el}\\
(\fc\el)(\alpha) =& \dim\gc{0,1,1\alpha,1,1}{\el}
\end{align}
the {\bf invariant functions} corresponding to two--dimensional
twisted cocycles of the adjoint representation of a Lie algebra
$\el$.

From Theorem~\ref{tvr1koho} follows immediately:
\begin{cor}\label{invarfunc2} If two complex Lie algebras $\el,\wt\el$ are
isomorphic, $\el \cong \widetilde{\el}$, then it holds:
\begin{enumerate}
\item $\fa\el = \fa\wt\el$,
\item $\fb\el = \fb\wt\el$,
\item $\fc\el = \fc\wt\el$.
\end{enumerate}
\end{cor}
\subsection{Invariant Functions $\fa$, $\fb$ and $\fc$ of Low--dimensional Lie Algebras}
We now investigate the behaviour of the functions $\fb,\,\fc$ in
dimensions three and four. It was shown in \cite{NHd} that the invariant function $\fa$
classifies indecomposable three--dimensional complex Lie algebras.
Moreover, observing the tables in Appendix A, the following theorem holds.
\begin{thm}[Classification of three--dimensional complex Lie
algebras]\label{class3dim}$\,$\newline
  Two three--dimensional complex Lie
  algebras $\el$, $\wt\el$ are isomorphic if and only if
$\fa\el=\fa\wt\el$.
\end{thm}

Observing the tables of $\fc$, we may also derive a quite interesting fact -- the function $\fc$
alone also classifies three--dimensional Lie algebras:

\begin{thm}[Classification of three--dimensional complex Lie
algebras II]\label{class3dim2}$\,$\newline
  Two three--dimensional complex Lie
  algebras $\el$ and $\wt\el$ are isomorphic if and only if $\fc\el=\fc\wt\el$.
\end{thm}

Theorem~\ref{class3dim} (or
\ref{class3dim2}) provided complete classification of
three--dimensional complex Lie algebras. We show in this section
that combined power of the functions $\fa$ and $\fb$ distinguishes
among all complex four--dimensional Lie algebras. We define the {\bf number of occurrences} of $j\in\Com$ in a complex function~$f$.
Let $j$ be in the range of values of $f$. If there exist only
finitely many mutually distinct numbers $x_1,\dots,x_m\in \Com$ for
which $f(x_1)=\dots=f(x_m)=j$ holds then we write $$f:j_m$$ and say
that $j$ {\bf occurs} in $f$ $m$--times; otherwise we write $f:j$.

\begin{thm}[Classification of four--dimensional complex Lie
algebras]\label{class4dim}  Two four--dimensional complex
Lie algebras $\el$ and $\wt\el$ are isomorphic if and only if
$\fa\el=\fa\wt\el$ and $\fb\el=\fb\wt\el$.
\end{thm}
\begin{proof}
See Appendix B. \end{proof}
An efficient algorithm for the identification of four--dimensional
Lie algebras was quite recently published in \cite{AY}. We may now formulate an alternative {\it
algorithm}: take a four--dimensional complex Lie algebra~$\el$ and
\begin{enumerate}
\item Calculate $\fa \el$ and $\fb\el$.
\item The range of values of the functions $\fa$ and $\fb$ and the
number of their occurrences determines the label (g-$k$),
$k=1,\dots,34$ in Appendix A.
\item The algebra is now identified up to the exact value of
parameter(s) of the parametric continuum. These parameters are
determined in the following cases:
\begin{itemize}
\item[(g-\ref{g34})]
Pick any of the two values $z\in\Com, z\neq 1$, which satisfy
$\fa\el(z)=6$, and put $a=z$. Then $\el\cong
\g_{3,4}(a)\oplus\g_1$ holds.
\item[(g-\ref{g42})]
 There are two different complex numbers $z_1,z_2\neq 0$ which satisfy $\fb\el(z_1)=\fb\el(z_2)=13$.
If $z_1-1=2/z_2$ holds then put $a=z_1-1$, otherwise put
$a=z_2-1$. Then $\el\cong \g_{4,2}(a)$ holds.
\item[(g-\ref{g45ab})]
There are three mutually different complex numbers
$z_1,z_2,z_3\neq 0,-1$ which satisfy
$\fb\el(z_1)=\fb\el(z_2)=\fb\el(z_3)=13$. Put
$a=\frac{z_3+1}{z_2+1}$, $b=\frac{z_2z_3-1}{z_2+1}$. Then
$\el\cong \g_{4,5}(a,b)$ holds.
\item[(g-\ref{g45am1ma})]
Pick any of the six values $z\in\Com$, which satisfy
$\fa\el(z)=5$, and put $a=z$. Then $\el\cong \g_{4,5}(a,-1-a)$
holds.
\item[(g-\ref{g45aas})]
Pick any of the two values $z\in\Com, z\neq 1$, which satisfy
$\fa\el(z)=6$, and put $a=z$. Then $\el\cong \g_{4,5}(a,a^2)$
holds.
\item[(g-\ref{g45a1})]
Take the value $z\in\Com$, which satisfies $\fb\el(z)=15$, and put
$a=z-1$. Then $\el\cong \g_{4,5}(a,1)$ holds.
\item[(g-\ref{g45am1})] Pick any of the two values $z\in\Com$, which satisfy
$\fb\el(z)=13$, and put $a=z+1$. Then $\el\cong \g_{4,5}(a,-1)$
holds.
\item[(g-\ref{g48})] Pick any of the two values $z\in\Com$, $z\neq 2$ which satisfy
$\fa\el(z)=4$, and put $a=z$. Then $\el\cong \g_{4,8}(a)$ holds.
\end{itemize}
\end{enumerate}

We demonstrate the above algorithm of identification on the
following example.
\begin{example}\label{AYE1}
In \cite{AY}, a four--dimensional algebra $\el_1$ was introduced:
\begin{center}
\begin{tabular}{clll}
  $\el_1:$ & $[e_1,e_2]=-e_1-e_2+e_3,$ & $[e_1,e_3]=-6e_2+4e_3,$& $[e_1,e_4]=2e_1-e_2+e_4,$ \\
          & $[e_2,e_3]=3e_1-9e_2+5e_3,$  & $[e_2,e_4]=4e_1-2e_2+2e_4,$  & $[e_3,e_4]=6e_1-3e_2+3e_4.$\\
\end{tabular}
\end{center}
\begin{enumerate}\item
Computing the functions $\fa\el_1$ and $\fb\el_1$ one obtains:
{\small
\begin{center}\vspace{-8pt}
\begin{tabular}[t]{|l||c|c|c|c|}
\hline \parbox[l][20pt][c]{0pt}{}   $\alpha$ & 1 & 2&
$\frac{1}{2}$& \\ \hline
\parbox[l][20pt][c]{0pt}{} $\fa\el_1(\alpha)$ & 6 & 5 & 5 & 4  \\ \hline
\end{tabular}\qquad
\begin{tabular}[t]{|l||c|c|c|}
\hline  \parbox[l][20pt][c]{0pt}{}  $\alpha$ & 3 & 1 &
\\ \hline
\parbox[l][20pt][c]{0pt}{} $\fb\el_1(\alpha)$ & 13  & 13 & 12\\
\hline
\end{tabular}
\end{center}}
\item The combination of occurrences $\fa\el_1: 6_1,5_2,4$ and $\fb\el_1:
13_2,12$ is unique for the case~(g-\ref{g42}).
\item Since for $z_1=3$, $z_2=1$ the equality $z_1-1=2/z_2$ holds,
one has $a=z_1-1=2$ and $\el_1\cong\g_{4,2}(2)$.
\end{enumerate}
\end{example}

\section{Contractions of Lie Algebras}\label{CHcon}
\subsection{Continuous Contractions of Lie Algebras}

Suppose we have an arbitrary Lie algebra $\el=(V,\,[\,,\,])$ and a
continuous mapping $U: (0,1\rangle \map GL(V)$, i.~e. $U(\ep)\in
GL(V), \: 0<\ep\leq 1$.  If the limit
\begin{equation}\label{contr}
[x , y]_0=\lim_{\ep \map 0+}U(\ep )^{-1}[U(\ep )x, U(\ep )y]
\end{equation}
exists for all $x,y\in V$ then we call the algebra
$\el_0=(V,[\,,\,]_0)$ a {\bf one--parametric continuous contraction}
(or simply a {\bf contraction}) of the algebra $\el$ and write
$\el\map \el_0$. We call the contraction $\el\map \el_0$ {\bf
proper} if $\el\ncong \el_0$. Contraction to the Abelian algebra is
always possible via $U(\ep)= \ep\, 1$.

It is well known that if $\el\map \el_0$ is any one--parametric
continuous contraction of a Lie algebra $\el$ then $\el_0$ is also
a Lie algebra. Invariant characteristics of Lie algebras change
after a contraction. The relation among these characteristics
before and after a contraction form useful necessary contraction
criteria. For example, such a set of these criteria, which
provided the complete classification of contractions of three and
four--dimensional Lie algebras, has been found in~\cite{Nes}. Our
aim is to state new necessary contraction criteria using
$(\alpha,\beta,\gamma)$--derivations and twisted cocycles.

\begin{thm}\label{dimderconthm}
Let $\el $ be a complex Lie algebra, $\el\map \el_0$ and $q\in
\N$. Then for any $(q+1)\times(q+1)$ complex symmetric matrix
$\kappa$
\begin{equation}\label{dimdercon2}
  \dim Z^{q} (\el,\ad_\el,\kappa) \leq \dim  Z^{q} (\el_0,\ad_{\el_0},\kappa)
\end{equation}
holds.
\end{thm}
\begin{proof}
Suppose that the contraction $\el\map \el_0$ is performed by the
mapping $U$, i.~e. $[x,y]_0=\lim_{\ep\map 0+}[x,y]_\ep ,$ where
$$[x,y]_\ep=U(\ep)^{-1}[U(\ep)x,U(\ep)y],\q \forall x,y\in\el.$$
Suppose $\el=(V,[\,,\,])$ and let us fix a basis
$\{x_1,\dots,x_n\}$ of $V$. We denote the structural constants of
the algebra $\el$ by~$c_{ij}^k$ and the structural constants of
the algebras $\el_\ep=(V,[\,,\,]_\ep)$ by~$c_{ij}^k(\ep)$. Then it
holds
\begin{equation}\label{limitstruc}
c_{ij}^k(0)=\lim_{\ep\map 0+}c_{ij}^k(\ep),
\end{equation}
where $c_{ij}^k(0)$ are the structural constants of $\el_0$.  The
dimension of the space $Z^{q} (\el,\ad_\el,\kappa)$ is determined
via the relation
\begin{equation}\label{rank}
  \dim Z^{q} (\el,\ad_\el,\kappa)= \dim
C^q(\el,\el)-\operatorname{rank} S^q(\el,\kappa),
\end{equation}
where $S^q(\el,\kappa)$ is the matrix corresponding to the linear
system of equations generated from (\ref{koho1t}). We write the
explicit form of this system for $q=1$. Then we obtain from
(\ref{koho1t}) that $D=(D_{ij})\in Z^{1}
\left(\el,\ad_\el,\left(\begin{smallmatrix}
  \beta &  \alpha \\
   \alpha & \gamma
\end{smallmatrix}\right) \right)$ if and only if the
linear system with the matrix $ S^{1}
\left(\el,\left(\begin{smallmatrix}
  \beta &  \alpha \\
   \alpha & \gamma
\end{smallmatrix}\right) \right) $ is satisfied
\begin{equation}\label{sys11}
 S^{1}
\left(\el,\left(\begin{smallmatrix}
  \beta &  \alpha \\
   \alpha & \gamma
\end{smallmatrix}\right) \right):\q \sum_{r=1}^n-\alpha c^r_{ij}D_{sr}+\beta
c^s_{rj}D_{ri}+\gamma c^s_{ir}D_{rj}=0, \q \forall i,j,s\in
\{1,\dots,n\},
\end{equation}
and similarly for $q>1$. Since $\el_\ep\cong\el$ holds for all
$0<\ep \leq 1 $, we see from Corollary~\ref{invdim} that
\begin{equation}\label{invzet}
\dim Z^{q} (\el,\ad_\el,\kappa)=\dim Z^{q}
(\el_\ep,\ad_{\el_\ep},\kappa), \q 0<\ep \leq 1,\,q\in\N.
\end{equation}
Since the relation
\begin{equation}\label{cql}
\dim C^q(\el,\el)= \dim C^q(\el_\ep,\el_\ep)= \dim
C^q(\el_0,\el_0),\q 0<\ep \leq 1,\,q\in\N
\end{equation}
holds, the relations~(\ref{rank}), (\ref{invzet}) then imply that
\begin{equation}\label{rank2}
\operatorname{rank} S^q(\el,\kappa)= \operatorname{rank}
S^q(\el_\ep,\kappa), \q 0<\ep \leq 1,\,q\in\N.
\end{equation}
The rank of the matrix $S^q(\el,\kappa)$ is equal to $r$ if and
only if there exists a non-zero minor of the order $r$ and every
minor of order higher than $r$ is zero. It follows from
(\ref{rank2}) that all minors of the orders higher than $r$ of the
matrices $S^q(\el_\ep,\kappa)$ are zeros. Since the equality
(\ref{limitstruc}) holds, all minors of the matrices
$S^q(\el_\ep,\kappa)$ converge to the minors of the matrix
$S^q(\el_0,\kappa)$. Thus, as the limits of zero functions, all
minors of order higher than $r$ of the matrix $S^q(\el_0,\kappa)$
are also zero. Therefore $\operatorname{rank}
S^q(\el_0,\kappa)\leq r$ and the statement of the theorem follows
from~(\ref{rank}) and~(\ref{cql}).
\end{proof}
There exist other necessary contraction criteria, similar to
(\ref{dimdercon2}) -- certain inequalities
between invariants. However, one very powerful criterion is quite
unique. This highly non-trivial theorem, very useful in
\cite{Bur2,Bur1,Nes}, was originally proved in \cite{Bor}.
\begin{thm}
If $\el_0$ is a proper contraction of a complex Lie algebra $\el$
then it holds:
\begin{equation}\label{dimdercon3}
  \dim \der \el < \dim \der \el_0.
\end{equation}
\end{thm}
\begin{cor}\label{concritmain}
If $\el_0$ is a proper contraction of a complex Lie algebra $\el$
then it holds: \begin{enumerate}\item $\fa\el\leq\fa\el_0 $ \item
$\fa\el(1)<\fa\el_0(1)$.
\end{enumerate}
\end{cor}
\begin{proof}
Since $\fa\el (\alpha)=\dim Z^{1}
\left(\el,\ad_\el,\left(\begin{smallmatrix}
  1 &  \alpha \\
   \alpha & 1
\end{smallmatrix}\right) \right)$ the first inequality follows from
(\ref{dimdercon2}) and the second from $\fa\el (1)= \dim \der \el$
and (\ref{dimdercon3}).
\end{proof}
\begin{cor}\label{concrit2}
If $\el_0$ is a contraction of a complex Lie algebra $\el$ then it
holds: \begin{enumerate}\item $\fb\el\leq\fb\el_0 $ \item
$\fc\el\leq\fc\el_0$.
\end{enumerate}
\end{cor}
\begin{proof}
Since $\fb\el (\alpha)=\dim Z^{2}
\left(\el,\ad_\el,\left(\begin{smallmatrix}
  \alpha &  1 & 1 \\
   1 & \alpha & 1 \\
   1 & 1 &\alpha
\end{smallmatrix}\right) \right)$ the first inequality follows from
(\ref{dimdercon2}); the proof of the second condition is
analogous.
\end{proof}

\subsection{Continuous Contractions of Low--dimensional Lie Algebras}

We have used the invariant function $\fa$ to classify all
three--dimensional Lie algebras in Theorem~\ref{class3dim}. We now
employ the necessary contraction criterion of
Corollary~\ref{concritmain} to describe all possible contractions
among these algebras. The behaviour of the function $\fa$ determines
the classification and contractions of three--dimensional Lie
algebras. Contractions of three--dimensional algebras were the most
recently classified in~\cite{Nes}:
\begin{thm}\label{contract3dim}
Only the following proper contractions among
three--dimensional Lie algebras exist:
\begin{enumerate}
\item $\g_{3,4}(-1)$ is a contraction of $\slp(2,\Com)$,
\item $\g_{3,3}$ is a contraction of $\g_{3,2}$,
\item $\g_{3,1}$ is a contraction of $\g_{3,2}$, $\g_{3,4}(a)$,
$\g_{2,1}\oplus\g_1$ and $\slp(2,\Com)$.

\item all algebras contract to the Abelian algebra
\end{enumerate}
\end{thm}
Analysis of all possible pairs of three--dimensional Lie
algebras leads us to the following theorem.
\begin{thm}[Contractions of three--dimensional complex Lie
algebras]\label{conthmmain}$\,$\newline
  Let $\el$, $\el_0$ be two three--dimensional complex Lie
  algebras. Then there exists a proper one--parametric continuous
  contraction $\el\map\el_0$ if and only if
  \begin{equation*}
  \fa\el\leq\fa\el_0 \quad \text{and}\quad \fa\el(1)<\fa\el_0(1).
\end{equation*}
\end{thm}
\begin{proof}
$\Rightarrow$ : This implication is, in fact,
Corollary~\ref{concritmain}.

$ \Leftarrow$: This implication follows from a direct comparison of
the tables of the invariant functions $\fa$ of three--dimensional
Lie algebras in Appendix A and Theorem~\ref{contract3dim}.
\end{proof}

We discuss the application of the criteria of the
Corollaries~\ref{concritmain} and \ref{concrit2} to the
four--dimensional Lie algebras in the following examples.
\begin{example}
To demonstrate behaviour of the functions $\fa,\fb$ and $\fc$ in
dimension four, we consider the following sequence of contractions
\cite{Bur1,Nes} : $$\slp(2,\Com) \oplus\g_1\ \map\ \g_{4,8}(-1)\
\map\ \g_{3,4}(-1)\oplus\g_1\ \map\ \g_{4,1}\ \map\
\g_{3,1}\oplus\g_1\ \map\ 4\g_1. $$ Note in Table~\ref{tab2}, how
the value of each invariant function is greater or equal than the
value in the previous row. As expected, the strict inequality for
the values~$\fa(1)$ holds -- in this case the sequence of
dimensions: $4,\,5\,,6\,,7\,,10\,,16$. The strict increase of
values is also identified in the following cases: $\fa(2)$,
'generic values' of $\fa$, $\fb(1/2)$ and 'generic values'
of~$\fb$. These conjectures of strict inequalities are, however,
not valid for the general case of a contraction in dimension four.
{\small
\begin{table}[!ht]
\begin{center}
\centering \caption[l]{\it Invariant functions $\fa,\, \fb$ and
$\fc$ of the contraction sequence: $\slp(2,\Com) \oplus\g_1\ \map\
\g_{4,8}(-1)\ \map\ \g_{3,4}(-1)\oplus\g_1\ \map\ \g_{4,1}\ \map\
\g_{3,1}\oplus\g_1\ \map\ 4\g_1. $} \label{tab2}\vspace{-6pt}
\begin{tabular}[t]{|c||c|c|c|c|c||c|c|c|c|c||c|c|c|c|}
\cline{1-15} \parbox[l][22pt][c]{0pt}{} \hspace{108pt}
&\multicolumn{5}{|c||}{$\fa (\al)$}& \multicolumn{5}{|c||}{$\fb
(\al)$} & \multicolumn{4}{c|}{$\fc (\al)$} \\

\hline
\parbox[l][22pt][c]{0pt}{}  $\al$ & -1& 0& 1 &2&  & -1 & 0 & 1 & $\frac{1}{2}$ &  & 0 & 1 & 2 &\\ \hline\hline

\parbox[l][22pt][c]{0pt}{}  $\slp(2,\Com)\oplus\g_1$ & 6& 4& 4 &2& 1 & 14 & 12 & 12 & 10 & 9  &  0 & 0 & 1 & 0\\ \hline

\parbox[l][22pt][c]{0pt}{}  $\g_{4,8}(-1)$ & 6& 4& 5 &4& 4 & 14 & 12 & 13 &12 &12  &  0 & 0& 1 &0\\ \hline

\parbox[l][22pt][c]{0pt}{}  $\g_{3,4}(-1)\oplus\g_1$ & 7& 7& 6 &5& 5 & 16 & 16 & 15 & 14 & 14 &  3 & 3 & 3 & 1\\ \hline

\parbox[l][22pt][c]{0pt}{}  $\g_{4,1}$ & 7& 7& 7 &7& 7 & 16 & 16 & 15 & 15 & 15 & 3& 3 & 3 & 3\\ \hline

\parbox[l][22pt][c]{0pt}{}  $\g_{3,1}\oplus\g_1$ & 10& 11& 10 &10& 10 & 19 & 20 & 19 & 19  &19  & 8& 11 & 8 & 8\\ \hline

\parbox[l][22pt][c]{0pt}{}  $4\g_1$ & 16& 16& 16 &16& 16 & 24 & 24 & 24 &24 & 24 & 24& 24 & 24 &24\\ \hline

\end{tabular}\end{center}
\end{table} }
\end{example}

\begin{example}
Consider the pair of parametric four--dimensional Lie algebras
$\g_{4,2}(a),\,a\neq 0,\pm 1,-2$ and $\g_{4,5}(a',1)$, $a'\neq 0,\pm
1,-2$. There are two possibilities, how the corresponding tables of
the invariant functions $\fb$, in Appendix~A cases~(g-\ref{g42}) and
(g-\ref{g45a1}), can satisfy $\fb\g_{4,2}(a)\leq \fb\g_{4,5}(a',1)$.
The first possibility leads to conditions $a'+1=2/a$ and $a+1=2/a'$
-- these have solutions $a=a'=1,-2$ and we excluded them. The second
possibility implies $a=a'$. The necessary condition 1. of
Corollary~\ref{concrit2} therefore admits only the contraction
$\g_{4,2}(a)\map\g_{4,5}(a,1)$. This contraction indeed exists
\cite{Nes}. In Table~\ref{tab3} we summarize the behaviour of the
functions $\fa,\, \fb$ and $\fc$. Note that the function $\fa$ grows
only at the points $1,\,a,\,\frac{1}{a}$ and the function $\fb$ only
at one(!) point $1+a$.
{\small
\begin{table}[h]
\begin{center}
\centering \caption[l]{\it Invariant functions $\fa,\, \fb$ and
$\fc$ of the contraction:
$\g_{4,2}(a)\map\g_{4,5}(a,1)$\hspace{50pt}}\label{tab3}\vspace{3pt}
\begin{tabular}[b]{|c||c|c|c|c||c|c|c||c|c|c|c|}
\hline\parbox[l][22pt][c]{0pt}{} \hspace{108pt}
&\multicolumn{4}{c||}{$\fa (\al)$}& \multicolumn{3}{|c||}{$\fb
(\al)$} & \multicolumn{4}{c|}{$\fc (\al)$} \\ \hline
\parbox[l][22pt][c]{0pt}{}  $\al$ & $\:\:1\:\:$ & $\:\:a\:\:$& $\:\:\frac{1}{a}\:\:$ &\hspace{16pt}  & $1+a$ & $\:\:\frac{2}{a}\:\:$ & \hspace{16pt}& $\:2\:$ & $1+a$ & $1+\frac{1}{a}$ &\hspace{12pt} \\ \hline\hline

\parbox[l][22pt][c]{0pt}{}  $\g_{4,2}(a)$ & 6& 5& 5 &4& 13 & 13 & 12 & 3 & 1 & 1  &  0 \\ \hline

\parbox[l][22pt][c]{0pt}{}  $\g_{4,5}(a,1)$ & 8& 6& 6 &4& 15 & 13 & 12 & 7 &2 &2  &  0 \\ \hline

\end{tabular}\end{center}
\end{table} }
\end{example}
\begin{example}\label{fbcon}
Consider the pair of four--dimensional Lie algebras $\g_{4,7}$ and
$\g_{4,2}(1)$. The necessary conditions $\fa\g_{4,7}\leq
\fa\g_{4,2}(1), $ $[\fa\g_{4,7}](1)< [\fa\g_{4,2}(1)](1) $ and
$\fb\g_{4,7}\leq \fb\g_{4,2}(1)$ are satisfied. But since it holds
$$1=[\fc\g_{4,7}]\left(\frac{3}{2}\right)>
[\fc\g_{4,2}(1)]\left(\frac{3}{2}\right)=0, $$ a contraction is
not possible.
\end{example}

  \subsection{Graded Contractions of Lie algebras}
Consider a graded
contraction of the Pauli graded $\slp(3,\Com)$ which was in~\cite{HN2} denoted by
$\ep^{17,7}(a)$, $a\neq 0$. We may determine this graded contraction by
listing its non--zero commutation relations in $\Z_3$--labeled
basis $( l_{01}, l_{02}, l_{10},  l_{20}, l_{11}, l_{22},
l_{12},l_{21})$: $$
\begin{array}{ll}
\el_{17,7}(a): \quad & [l_{01},l_{10}]=-al_{11},\
[l_{01},l_{20}]=l_{21},\ [l_{01},l_{11}]=l_{12},\
[l_{01},l_{22}]=l_{20},\\
  & [l_{02},l_{10}]=l_{12},\ [l_{02},l_{22}]=l_{21},\ [l_{10},l_{11}]=l_{21},\q a\neq 0 .\\
\end{array} $$ Lie algebras $\el_{17,7}(a)$ are all
indecomposable, nilpotent and their derived series, lower central series, upper central series and the number of formal Casimir invariants~\cite{inv} coincide.
The invariant function $\fa$ has the following form:
{\small
\begin{center}
\begin{tabular}[t]{|l||c|c|c|}
\hline \parbox[l][20pt][c]{0pt}{}   $\alpha$  & 0  & 1 & \\ \hline
\parbox[l][20pt][c]{0pt}{} $\fa\el_{17,7}(a)(\alpha)$ & 20 & 19 & 18\\ \hline
\end{tabular}
\end{center}}
In this case, the function $\fa$ completely fails  -- does not
depend on $a\neq 0$. We are able, however, to advance by
calculation of the function $\fb$:
{\small
\begin{center}
\begin{tabular}[t]{|l||c|c|c|c|}
\multicolumn{5}{l}{$a=1$} \\ \hline \parbox[l][20pt][c]{0pt}{}
$\alpha$  & 0 & 1 & -1 &\\ \hline \parbox[l][20pt][c]{0pt}{}
$\fb\el_{17,7}(1)(\alpha)$ & 112 & 83 & 81 & 80\\ \hline
\end{tabular}\qquad
\begin{tabular}[t]{|l||c|c|c|c|}
\multicolumn{5}{l}{$a=-1$} \\ \hline \parbox[l][20pt][c]{0pt}{}
$\alpha$  & 0 & 1 & -1 &\\ \hline \parbox[l][20pt][c]{0pt}{}
$\fb\el_{17,7}(-1)(\alpha)$ & 104 & 83 & 81 & 80\\ \hline
\end{tabular}

\vspace{4pt}

\begin{tabular}[t]{|l||c|c|c|c|}
\multicolumn{5}{l}{$a=\frac{1}{4}+\frac{\sqrt{7}}{4}i$}\parbox[l][20pt][c]{0pt}{}
\\ \hline
\parbox[l][26pt][c]{0pt}{} $\alpha$  & 0 & 1 & $-\frac{1}{4}-\frac{\sqrt{7}}{4}i$ &\\ \hline
\parbox[l][26pt][c]{0pt}{} $\fb\el_{17,7}\left(\frac{1}{4}+\frac{\sqrt{7}}{4}i\right)(\alpha)$ & 104 & 82 &
82 & 80\\ \hline
\end{tabular}

\vspace{4pt}

\begin{tabular}[t]{|l||c|c|c|c|}
\multicolumn{5}{l}{$a=\frac{1}{4}-\frac{\sqrt{7}}{4}i$}\parbox[l][20pt][c]{0pt}{}
\\ \hline
\parbox[l][26pt][c]{0pt}{} $\alpha$  & 0 & 1 & $-\frac{1}{4}+\frac{\sqrt{7}}{4}i$ &\\ \hline
\parbox[l][26pt][c]{0pt}{} $\fb\el_{17,7}\left(\frac{1}{4}-\frac{\sqrt{7}}{4}i\right)(\alpha)$ & 104 & 82 &
82 & 80\\ \hline
\end{tabular}

\vspace{4pt}

\begin{tabular}[t]{|l||c|c|c|c|}
\multicolumn{5}{l}{$a=\frac{1}{3}$}\parbox[l][20pt][c]{0pt}{}
\\ \hline
\parbox[l][26pt][c]{0pt}{} $\alpha$  & 0 & 1 & $-\frac{1}{3}$ &\\ \hline
\parbox[l][26pt][c]{0pt}{} $\fb\el_{17,7}\left(\frac{1}{3}\right)(\alpha)$ & 104 & 83 &
81 & 80\\ \hline
\end{tabular}

\vspace{4pt}

\begin{tabular}[t]{|l||c|c|c|c|c|}
\multicolumn{6}{l}{$a \neq
0,\pm1,\frac{1}{3},\frac{1}{4}\pm\frac{\sqrt{7}}{4}i
$}\parbox[l][20pt][c]{0pt}{} \\ \hline
\parbox[l][20pt][c]{0pt}{}   $\alpha$  & 0 & 1 & $-a$ & $-\frac{1}{2}+\frac{1}{2a}$ & \\
  \hline \parbox[l][20pt][c]{0pt}{}
$\fb\el_{17,7}(a)(\alpha)$ & 104 & 82 & 81 & 81 & 80 \\ \hline
\end{tabular}
\end{center} }
In order to verify that
\begin{equation}\label{equ177}
\fb\el_{17,7}(a):\,  104_1,82_1,81_2,80, \q a \neq
0,\pm1,\frac{1}{3},\frac{1}{4}\pm\frac{\sqrt{7}}{4}i
\end{equation}
we have to check the equality
\begin{equation*}
-a=-\frac{1}{2}+\frac{1}{2a}
\end{equation*}
which has the solutions $\frac{1}{4}\pm\frac{\sqrt{7}}{4}i$. Thus,
(\ref{equ177}) is verified.

We proceed to solve the relation
$$\fb\el_{17,7}(a)=\fb\el_{17,7}(a'), \q a,a' \neq
0,\pm1,\frac{1}{3},\frac{1}{4}\pm\frac{\sqrt{7}}{4}i $$ and we
obtain
\begin{enumerate}
\item If $-a=-a',\,
-\frac{1}{2}+\frac{1}{2a}=\frac{1}{2}+\frac{1}{2a'}$ then $a=a'$.
\item If $-a=-\frac{1}{2}+\frac{1}{2a'},\,-a'=-\frac{1}{2}+\frac{1}{2a}
$ then $a=a'=\frac{1}{4}\pm\frac{\sqrt{7}}{4}i$.
\end{enumerate}
The second case is not possible. Observing that all other tables
of the function $$\fb\el_{17,7}(a),\, a\neq
\pm1,\frac{1}{3},\frac{1}{4}\pm\frac{\sqrt{7}}{4}i $$ are mutually
different, we have: if $\fb\el_{17,7}(a)=\fb\el_{17,7}(a'), \,
a,a' \neq 0$ then $a=a'$. We conclude that even though the
function $\fa$ did not distinguish the algebras in nilpotent
parametric continuum, the function $\fb$ provided their complete
description.

  \section{Concluding Remarks}
\noindent
\begin{itemize}
\item
The invariant functions $\fa$ and $\fb$ are able to classify all
four--dimensional Lie algebras and provide also necessary contraction criteria. In order to obtain stronger
contraction criteria, we also defined the function $\fc$ -- a
supplement to the functions $\fa$ and $\fb$ (see
Example~\ref{fbcon}). However, the combined forces of the
Corollaries~\ref{concritmain} and \ref{concrit2} do
not provide us with a complete classification of contractions of
four--dimensional Lie algebras.
\item Existence of invariant functions, arising from the
concept of two--dimensional twisted cocycles and allowing classification of continuous contractions of four--dimensional complex Lie algebras, remains an open problem. The complete description of the spaces of
two--dimensional twisted cocycles for four--dimensional Lie algebras
would solve the existence of such functions explicitly. However,
such a complete description seems, at the moment, out of reach.
\item The contraction criterion formulated in Theorem 4.1 is a natural generalization of contraction criteria which involve standard cohomology cocycles -- see e.g.~\cite{Bur2,Campocoh}.
\item The invariant function $\fa$ can be easily generalized and used as an invariant of an arbitrary anti-commutative or commutative algebra -- it can, for example, describe all two--dimensional complex Jordan algebras and their contractions~\cite{Hri}.
\item In contrast to algebraic approach of the 'trace' invariants $C_{pq}$ and $\chi_i$~\cite{Bur1,AY}, the concept of invariant cocycles provides a new knowledge about the dimensions of highly non--trivial structures, interlaced
with given Lie algebra. Considering
parametric continua of nilpotent algebras, which firstly appear in dimension $7$, the invariant functions
$\fa,\fb,\fc$ seem to be even more important. In these cases, the
behaviour of these invariant functions
 is quite unique and irreplaceable.
\end{itemize}

\section*{Acknowledgements}
The authors are grateful to J.~Tolar for numerous stimulating discussions. Partial support by the Ministry of Education of Czech Republic (projects MSM6840770039 and LC06002) is gratefully acknowledged. 

\section*{ Appendix A}
Appendix A contains the classification of complex Lie algebras up to
dimension four and the invariant functions $\fa,\fb,\fc$. We
basically follow the notation of~\cite{Nes}.
Instead of the symbols $\fa\el,\fb\el,\fc\el$, abbreviated symbols
$\fa,\fb,\fc$ are used. Blank spaces in the tables of the
functions $\fa,\fb,\fc$ denote general complex numbers, different
from all previously listed values, e. g. it holds:
$$\fa\g_{3,4}(-1)(\al)=3,\q\alpha\in \Com,\,\al\neq\pm 1.$$
  {\small
\subsection*{Two--dimensional Complex Lie Algebras}
\hspace{16pt}\newline
\begin{tabular}{ll}
\hspace{12pt}$2\g_{1}: \quad$ & Abelian
\end{tabular}
\begin{center}
\begin{tabular}[t]{|l||c|c|}
\hline \parbox[l][20pt][c]{0pt}{}   $\alpha$ & 1 &  \\ \hline
\parbox[l][20pt][c]{0pt}{} $\fa(\alpha)$ & 4 & 4  \\ \hline
\end{tabular}\qquad
\begin{tabular}[t]{|l||c|c|c|}
\hline  \parbox[l][20pt][c]{0pt}{}  $\alpha$ &   \\ \hline
\parbox[l][20pt][c]{0pt}{} $\fb(\alpha)$ & 2\\
\hline
\end{tabular}\qquad
\begin{tabular}[t]{|l||c|c|c|}
\hline  \parbox[l][20pt][c]{0pt}{}  $\alpha$ &   \\ \hline
\parbox[l][20pt][c]{0pt}{} $\fc(\alpha)$ & 2\\
\hline
\end{tabular}
\end{center}

\begin{tabular}{ll}
$\g_{2,1}: \quad$ & $ [e_1,e_2]=e_1$
\end{tabular}

\begin{center}
\begin{tabular}[t]{|l||c|c|c|}
\hline \parbox[l][20pt][c]{0pt}{}   $\alpha$ & 1 & 0 &  \\ \hline
\parbox[l][20pt][c]{0pt}{} $\fa(\alpha)$ & 2 & 3& 2  \\ \hline
\end{tabular}\qquad
\begin{tabular}[t]{|l||c|c|c|}
\hline  \parbox[l][20pt][c]{0pt}{}  $\alpha$ &   \\ \hline
\parbox[l][20pt][c]{0pt}{} $\fb(\alpha)$ & 2\\
\hline
\end{tabular}\qquad
\begin{tabular}[t]{|l||c|c|c|}
\hline  \parbox[l][20pt][c]{0pt}{}  $\alpha$ & 2 &  \\ \hline
\parbox[l][20pt][c]{0pt}{} $\fc(\alpha)$ & 1& 0\\
\hline
\end{tabular}
\end{center}

\subsection*{Three--dimensional Complex Lie Algebras}
\hspace{16pt}\newline
\begin{tabular}{ll}
\hspace{12pt}$3\g_{1}: \quad$ & Abelian
\end{tabular}
\begin{center}
\begin{tabular}[t]{|l||c|c|}
\hline \parbox[l][20pt][c]{0pt}{}   $\alpha$ & 1 &  \\ \hline
\parbox[l][20pt][c]{0pt}{} $\fa(\alpha)$ & 9 & 9  \\ \hline
\end{tabular}\qquad
\begin{tabular}[t]{|l||c|c|c|}
\hline  \parbox[l][20pt][c]{0pt}{}  $\alpha$ &   \\ \hline
\parbox[l][20pt][c]{0pt}{} $\fb(\alpha)$ & 9\\
\hline
\end{tabular}\qquad
\begin{tabular}[t]{|l||c|c|c|}
\hline  \parbox[l][20pt][c]{0pt}{}  $\alpha$ &   \\ \hline
\parbox[l][20pt][c]{0pt}{} $\fc(\alpha)$ & 9\\
\hline
\end{tabular}
\end{center}

\begin{tabular}{ll}
$\g_{2,1}\oplus \g_1: \quad$ & $ [e_1,e_2]=e_2$
\end{tabular}

\begin{center}
\begin{tabular}[t]{|l||c|c|c|}
\hline \parbox[l][20pt][c]{0pt}{}   $\alpha$ & 1 & 0 &  \\ \hline
\parbox[l][20pt][c]{0pt}{} $\fa(\alpha)$ & 4 & 6 & 4  \\ \hline
\end{tabular}\qquad
\begin{tabular}[t]{|l||c|c|c|}
\hline  \parbox[l][20pt][c]{0pt}{}  $\alpha$ &   \\ \hline
\parbox[l][20pt][c]{0pt}{} $\fb(\alpha)$ & 6\\
\hline
\end{tabular}\qquad
\begin{tabular}[t]{|l||c|c|c|c|}
\hline  \parbox[l][20pt][c]{0pt}{}  $\alpha$ & 1& 2&  \\ \hline
\parbox[l][20pt][c]{0pt}{} $\fc(\alpha)$ &2&2&1\\
\hline
\end{tabular}
\end{center}

\begin{tabular}{ll} $\g_{3,1}: \quad$ &$[e_2,e_3]=e_1$
\end{tabular}
\begin{center}
\begin{tabular}[t]{|l||c|c|c|}
\hline \parbox[l][20pt][c]{0pt}{}   $\alpha$ & 1 &   \\ \hline
\parbox[l][20pt][c]{0pt}{} $\fa(\alpha)$ & 6 & 6   \\ \hline
\end{tabular}\qquad
\begin{tabular}[t]{|l||c|c|c|}
\hline  \parbox[l][20pt][c]{0pt}{}  $\alpha$ & 0&  \\ \hline
\parbox[l][20pt][c]{0pt}{} $\fb(\alpha)$ &9& 8\\
\hline
\end{tabular}\qquad
\begin{tabular}[t]{|l||c|c|c|c|}
\hline  \parbox[l][20pt][c]{0pt}{}  $\alpha$ &   \\ \hline
\parbox[l][20pt][c]{0pt}{} $\fc(\alpha)$ &3\\
\hline
\end{tabular}
\end{center}

\begin{tabular}{ll} $\g_{3,2}: \quad$ &$[e_1,e_3]=e_1,\
[e_2,e_3]=e_1+e_2$
\end{tabular}
\begin{center}
\begin{tabular}[t]{|l||c|c|c|}
\hline \parbox[l][20pt][c]{0pt}{}   $\alpha$ & 1 &   \\ \hline
\parbox[l][20pt][c]{0pt}{} $\fa(\alpha)$ & 4 & 3   \\ \hline
\end{tabular}\qquad
\begin{tabular}[t]{|l||c|c|c|}
\hline  \parbox[l][20pt][c]{0pt}{}  $\alpha$ &  \\ \hline
\parbox[l][20pt][c]{0pt}{} $\fb(\alpha)$ &6\\
\hline
\end{tabular}\qquad
\begin{tabular}[t]{|l||c|c|c|c|}
\hline  \parbox[l][20pt][c]{0pt}{}  $\alpha$ & 2 &  \\ \hline
\parbox[l][20pt][c]{0pt}{} $\fc(\alpha)$ &2&0\\
\hline
\end{tabular}
\end{center}

\begin{tabular}{ll} $\g_{3,3}: \quad$ &$[e_1,e_3]=e_1,\
[e_2,e_3]=e_2$
\end{tabular}
\begin{center}
\begin{tabular}[t]{|l||c|c|c|}
\hline \parbox[l][20pt][c]{0pt}{}   $\alpha$ & 1 &   \\ \hline
\parbox[l][20pt][c]{0pt}{} $\fa(\alpha)$ & 6 & 3   \\ \hline
\end{tabular}\qquad
\begin{tabular}[t]{|l||c|c|c|}
\hline  \parbox[l][20pt][c]{0pt}{}  $\alpha$ &   \\ \hline
\parbox[l][20pt][c]{0pt}{} $\fb(\alpha)$ &6\\
\hline
\end{tabular}\qquad
\begin{tabular}[t]{|l||c|c|c|c|}
\hline  \parbox[l][20pt][c]{0pt}{}  $\alpha$ & 2&   \\ \hline
\parbox[l][20pt][c]{0pt}{} $\fc(\alpha)$ &6& 0\\
\hline
\end{tabular}
\end{center}

\begin{tabular}{ll} $\g_{3,4}(-1): \quad$ &$[e_1,e_3]=e_1,\
[e_2,e_3]=-e_2$
\end{tabular}
\begin{center}
\begin{tabular}[t]{|l||c|c|c|}
\hline \parbox[l][20pt][c]{0pt}{}   $\alpha$ & 1 & -1 &   \\
\hline
\parbox[l][20pt][c]{0pt}{} $\fa(\alpha)$ & 4 & 5 &3   \\ \hline
\end{tabular}\qquad
\begin{tabular}[t]{|l||c|c|c|}
\hline  \parbox[l][20pt][c]{0pt}{}  $\alpha$ & 0&  \\ \hline
\parbox[l][20pt][c]{0pt}{} $\fb(\alpha)$ &9& 7\\
\hline
\end{tabular}\qquad
\begin{tabular}[t]{|l||c|c|c|c|}
\hline  \parbox[l][20pt][c]{0pt}{}  $\alpha$ & 0&2&  \\ \hline
\parbox[l][20pt][c]{0pt}{} $\fc(\alpha)$ &2&2&0\\
\hline
\end{tabular}
\end{center}

\begin{tabular}{ll} $\g_{3,4}(a): \quad$ &$[e_1,e_3]=e_1,\
[e_2,e_3]=ae_2,\ a\neq 0,\pm 1$
\end{tabular}
\begin{center}
\begin{tabular}[t]{|l||c|c|c|c|}
\hline \parbox[l][20pt][c]{0pt}{}   $\alpha$ & 1 & $a$
&$\frac{1}{a}$&
\\ \hline
\parbox[l][20pt][c]{0pt}{} $\fa(\alpha)$ & 4 & 4& 4&3   \\ \hline
\end{tabular}\qquad
\begin{tabular}[t]{|l||c|c|c|}
\hline  \parbox[l][20pt][c]{0pt}{}  $\alpha$ &   \\ \hline
\parbox[l][20pt][c]{0pt}{} $\fb(\alpha)$ &6\\
\hline
\end{tabular}\qquad
\begin{tabular}[t]{|l||c|c|c|c|}
\hline  \parbox[l][20pt][c]{0pt}{}  $\alpha$ & 2&$1+a$ &
$1+\frac{1}{a}$&
\\ \hline
\parbox[l][20pt][c]{0pt}{} $\fc(\alpha)$ &2&1&1&0\\
\hline
\end{tabular}
\end{center}

\begin{tabular}{ll}$\slp (2,\Com): \quad$ &$[e_1,e_2]=e_1,\
[e_2,e_3]=e_3,\ [e_1,e_3]=2e_2$
\end{tabular}
\begin{center}
\begin{tabular}[t]{|l||c|c|c|c|}
\hline \parbox[l][20pt][c]{0pt}{}   $\alpha$ & 1 & $-1$& 2&   \\
\hline
\parbox[l][20pt][c]{0pt}{} $\fa(\alpha)$ & 3 & 5& 1&0   \\ \hline
\end{tabular}\qquad
\begin{tabular}[t]{|l||c|c|c|}
\hline  \parbox[l][20pt][c]{0pt}{}  $\alpha$ & 0&  \\ \hline
\parbox[l][20pt][c]{0pt}{} $\fb(\alpha)$ &9& 6\\
\hline
\end{tabular}\qquad
\begin{tabular}[t]{|l||c|c|c|c|}
\hline  \parbox[l][20pt][c]{0pt}{}  $\alpha$ & 2&   \\ \hline
\parbox[l][20pt][c]{0pt}{} $\fc(\alpha)$ &1&0\\
\hline
\end{tabular}
\end{center}
\subsection*{Four--dimensional Complex Lie Algebras}
\begin{enumerate}[(g-1)]
\item \begin{tabular}{ll} $4\g_{1}: \quad$ & Abelian
\end{tabular}
\vspace{-8pt}
\begin{center}
\begin{tabular}[t]{|l||c|c|}
\hline \parbox[l][20pt][c]{0pt}{}   $\alpha$ & 1 &  \\ \hline
\parbox[l][20pt][c]{0pt}{} $\fa(\alpha)$ & 16 & 16  \\ \hline
\end{tabular}\qquad
\begin{tabular}[t]{|l||c|c|c|}
\hline  \parbox[l][20pt][c]{0pt}{}  $\alpha$ &   \\ \hline
\parbox[l][20pt][c]{0pt}{} $\fb(\alpha)$ & 24\\
\hline
\end{tabular}\qquad
\begin{tabular}[t]{|l||c|c|c|}
\hline  \parbox[l][20pt][c]{0pt}{}  $\alpha$ &   \\ \hline
\parbox[l][20pt][c]{0pt}{} $\fc(\alpha)$ & 24\\
\hline
\end{tabular}
\end{center}
\item \begin{tabular}{ll} $\g_{2,1}\oplus 2\g_1: \quad$ & $
[e_1,e_2]=e_1$
\end{tabular}
\vspace{-8pt}\begin{center}
\begin{tabular}[t]{|l||c|c|c|c|}
\hline \parbox[l][20pt][c]{0pt}{}   $\alpha$ & 1 & $0$&
\\ \hline
\parbox[l][20pt][c]{0pt}{} $\fa(\alpha)$ & 8 & 11 & 8  \\ \hline
\end{tabular}\qquad
\begin{tabular}[t]{|l||c|c|c|}
\hline  \parbox[l][20pt][c]{0pt}{}  $\alpha$ & $0$  &
\\ \hline
\parbox[l][20pt][c]{0pt}{} $\fb(\alpha)$ & 16  & 14 \\
\hline
\end{tabular}\qquad
\begin{tabular}[t]{|l||c|c|c|c|}
\hline \parbox[l][20pt][c]{0pt}{}   $\alpha$ & 1 & $2$ &
\\ \hline
\parbox[l][20pt][c]{0pt}{} $\fc(\alpha)$ & 8 & 7 & 6 \\ \hline
\end{tabular}
\end{center}
\item\label{g21og21}\begin{tabular}{ll} $\g_{2,1}\oplus \g_{2,1}: \quad$ & $
[e_1,e_2]=e_1,\ [e_3,e_4]=e_3$
\end{tabular}
\vspace{-8pt}\begin{center}
\begin{tabular}[t]{|l||c|c|c|c|}
\hline \parbox[l][20pt][c]{0pt}{}   $\alpha$ & 1 & $0$&
\\ \hline
\parbox[l][20pt][c]{0pt}{} $\fa(\alpha)$ & 4 & 6 & 4  \\ \hline
\end{tabular}\qquad
\begin{tabular}[t]{|l||c|c|c|}
\hline  \parbox[l][20pt][c]{0pt}{}  $\alpha$ & $0$  & 1&
\\ \hline
\parbox[l][20pt][c]{0pt}{} $\fb(\alpha)$ & 12  & 12 &10 \\
\hline
\end{tabular}\qquad
\begin{tabular}[t]{|l||c|c|c|c|}
\hline \parbox[l][20pt][c]{0pt}{}   $\alpha$ & 1 & $2$ &
\\ \hline
\parbox[l][20pt][c]{0pt}{} $\fc(\alpha)$ & 2 & 2 & 0 \\ \hline
\end{tabular}
\end{center}

\item\begin{tabular}{ll} $\g_{3,1}\oplus \g_1: \quad$ &$[e_2,e_3]=e_1$
\end{tabular}
\vspace{-8pt}\begin{center}
\begin{tabular}[t]{|l||c|c|c|c|}
\hline \parbox[l][20pt][c]{0pt}{}   $\alpha$ & 1 & $0$&
\\ \hline
\parbox[l][20pt][c]{0pt}{} $\fa(\alpha)$ & 10 & 11 & 10  \\ \hline
\end{tabular}\qquad
\begin{tabular}[t]{|l||c|c|c|}
\hline  \parbox[l][20pt][c]{0pt}{}  $\alpha$ & $0$  &
\\ \hline
\parbox[l][20pt][c]{0pt}{} $\fb(\alpha)$ & 20  & 19 \\
\hline
\end{tabular}\qquad
\begin{tabular}[t]{|l||c|c|c|c|}
\hline \parbox[l][20pt][c]{0pt}{}   $\alpha$ & 1 &
\\ \hline
\parbox[l][20pt][c]{0pt}{} $\fc(\alpha)$ & 11 & 8 \\ \hline
\end{tabular}
\end{center}

\item\begin{tabular}{ll} $\g_{3,2}\oplus \g_1: \quad$
&$[e_1,e_3]=e_1,\ [e_2,e_3]=e_1+e_2$
\end{tabular}
\vspace{-8pt}\begin{center}
\begin{tabular}[t]{|l||c|c|c|c|}
\hline \parbox[l][20pt][c]{0pt}{}   $\alpha$ & 1 & $0$&
\\ \hline
\parbox[l][20pt][c]{0pt}{} $\fa(\alpha)$ & 6 & 7 & 5  \\ \hline
\end{tabular}\qquad
\begin{tabular}[t]{|l||c|c|c|}
\hline  \parbox[l][20pt][c]{0pt}{}  $\alpha$ & $1$  &
\\ \hline
\parbox[l][20pt][c]{0pt}{} $\fb(\alpha)$ & 13  & 12 \\
\hline
\end{tabular}\qquad
\begin{tabular}[t]{|l||c|c|c|c|}
\hline \parbox[l][20pt][c]{0pt}{}   $\alpha$ & 1 & $2$ &
\\ \hline
\parbox[l][20pt][c]{0pt}{} $\fc(\alpha)$ & 3 & 3 & 1 \\ \hline
\end{tabular}
\end{center}

\item\begin{tabular}{ll} $\g_{3,3}\oplus\g_1: \quad$
&$[e_1,e_3]=e_1,\ [e_2,e_3]=e_2$
\end{tabular}
\vspace{-8pt}\begin{center}
\begin{tabular}[t]{|l||c|c|c|c|}
\hline \parbox[l][20pt][c]{0pt}{}   $\alpha$ & 1 & $0$&
\\ \hline
\parbox[l][20pt][c]{0pt}{} $\fa(\alpha)$ & 8 & 7 & 5  \\ \hline
\end{tabular}\qquad
\begin{tabular}[t]{|l||c|c|c|}
\hline  \parbox[l][20pt][c]{0pt}{}  $\alpha$ & $1$  &
\\ \hline
\parbox[l][20pt][c]{0pt}{} $\fb(\alpha)$ & 15  & 12 \\
\hline
\end{tabular}\qquad
\begin{tabular}[t]{|l||c|c|c|c|}
\hline \parbox[l][20pt][c]{0pt}{}   $\alpha$ & 1 & $2$ &
\\ \hline
\parbox[l][20pt][c]{0pt}{} $\fc(\alpha)$ & 3 & 7 & 1 \\ \hline
\end{tabular}
\end{center}

\item\begin{tabular}{ll} $\g_{3,4}(-1)\oplus\g_1: \quad$
&$[e_1,e_3]=e_1,\ [e_2,e_3]=-e_2$
\end{tabular}
\vspace{-8pt}\begin{center}
\begin{tabular}[t]{|l||c|c|c|c|c|}
\hline \parbox[l][20pt][c]{0pt}{}   $\alpha$ & 1 & $0$ & $-1$&
\\ \hline
\parbox[l][20pt][c]{0pt}{} $\fa(\alpha)$ & 6 & 7 &7& 5  \\ \hline
\end{tabular}\qquad
\begin{tabular}[t]{|l||c|c|c|c|}
\hline  \parbox[l][20pt][c]{0pt}{}  $\alpha$ & $1$& 0 & $-1$&
\\ \hline
\parbox[l][20pt][c]{0pt}{} $\fb(\alpha)$ & 15  & 16&16&14 \\
\hline
\end{tabular}\qquad
\begin{tabular}[t]{|l||c|c|c|c|c|}
\hline \parbox[l][20pt][c]{0pt}{}   $\alpha$ & 1 & $2$ & 0&
\\ \hline
\parbox[l][20pt][c]{0pt}{} $\fc(\alpha)$ & 3 & 3 & 3&  1 \\ \hline
\end{tabular}
\end{center}

\item\label{g34}\begin{tabular}{ll} $\g_{3,4}(a)\oplus \g_1: \quad$
&$[e_1,e_3]=e_1,\ [e_2,e_3]=ae_2,\ a\neq 0,\pm 1$
\end{tabular}
\vspace{-8pt}\begin{center}
\begin{tabular}[t]{|l||c|c|c|c|c|c|}
\hline \parbox[l][20pt][c]{0pt}{}   $\alpha$ & 1 & $0$ &
$a$&$\frac{1}{a}$ &
\\ \hline
\parbox[l][20pt][c]{0pt}{} $\fa(\alpha)$ & 6 & 7 &6& 6& 5  \\ \hline
\end{tabular}\qquad
\begin{tabular}[t]{|l||c|c|c|c|}
\hline  \parbox[l][20pt][c]{0pt}{}  $\alpha$ & $1$& $a$ &
$\frac{1}{a}$&
\\ \hline
\parbox[l][20pt][c]{0pt}{} $\fb(\alpha)$ & 13  & 13&13&12 \\
\hline
\end{tabular}\qquad

\vspace{12pt}

\begin{tabular}[t]{|l||c|c|c|c|c|}
\hline \parbox[l][20pt][c]{0pt}{}   $\alpha$ & 1 & $2$ &
$1+a$&$1+\frac{1}{a}$ &
\\ \hline
\parbox[l][20pt][c]{0pt}{} $\fc(\alpha)$ & 3 & 3 & 2 &2&  1 \\ \hline
\end{tabular}
\end{center}

\item\begin{tabular}{ll}$\slp (2,\Com)\oplus\g_1: \quad$
&$[e_1,e_2]=e_1,\ [e_2,e_3]=e_3,\ [e_1,e_3]=2e_2$
\end{tabular}
\vspace{-8pt}\begin{center}
\begin{tabular}[t]{|l||c|c|c|c|c|}
\hline \parbox[l][20pt][c]{0pt}{}   $\alpha$ & 1 & $0$ & $-1$& 2&
\\ \hline
\parbox[l][20pt][c]{0pt}{} $\fa(\alpha)$ & 4 & 4 &6& 2 & 1  \\ \hline
\end{tabular}\qquad
\begin{tabular}[t]{|l||c|c|c|c|c|}
\hline  \parbox[l][20pt][c]{0pt}{}  $\alpha$ & $1$& 0 & $-1$&
$\frac{1}{2}$ &
\\ \hline
\parbox[l][20pt][c]{0pt}{} $\fb(\alpha)$ & 12  & 12&14&10&9 \\
\hline
\end{tabular}\qquad
\begin{tabular}[t]{|l||c|c|c|c|c|}
\hline \parbox[l][20pt][c]{0pt}{}   $\alpha$ & 2 &
\\ \hline
\parbox[l][20pt][c]{0pt}{} $\fc(\alpha)$ & 1 & 0 \\ \hline
\end{tabular}
\end{center}

\item\begin{tabular}{ll}$\g_{4,1}: \quad$ &$[e_2,e_4]=e_1,\
[e_3,e_4]=e_2$
\end{tabular}
\vspace{-8pt}\begin{center}
\begin{tabular}[t]{|l||c|c|}
\hline \parbox[l][20pt][c]{0pt}{}   $\alpha$ & 1 &  \\ \hline
\parbox[l][20pt][c]{0pt}{} $\fa(\alpha)$ & 7 & 7  \\ \hline
\end{tabular}\qquad
\begin{tabular}[t]{|l||c|c|c|}
\hline  \parbox[l][20pt][c]{0pt}{}  $\alpha$ & -1 & 0 & \\ \hline
\parbox[l][20pt][c]{0pt}{} $\fb(\alpha)$ & 16  & 16 & 15\\
\hline
\end{tabular}\qquad
\begin{tabular}[t]{|l||c|c|c|}
\hline \parbox[l][20pt][c]{0pt}{}   $\alpha$       &  \\ \hline
\parbox[l][20pt][c]{0pt}{} $\fc(\alpha)$ & 3 \\ \hline
\end{tabular}
\end{center}

\item\label{g42}\begin{tabular}{ll}$\g_{4,2}(a): \quad$ &$[e_1,e_4]=ae_1,\
[e_2,e_4]=e_2, \ [e_3,e_4]=e_2+e_3,\ a\neq 0,\pm 1, -2 $
\end{tabular}
\vspace{-8pt}\begin{center}
\begin{tabular}[t]{|l||c|c|c|c|}
\hline \parbox[l][20pt][c]{0pt}{}   $\alpha$ & 1 & $a$&
$\frac{1}{a}$& \\ \hline
\parbox[l][20pt][c]{0pt}{} $\fa(\alpha)$ & 6 & 5 & 5 & 4  \\ \hline
\end{tabular}\qquad
\begin{tabular}[t]{|l||c|c|c|}
\hline  \parbox[l][20pt][c]{0pt}{}  $\alpha$ & $1+a$ &
$\frac{2}{a}$ &
\\ \hline
\parbox[l][20pt][c]{0pt}{} $\fb(\alpha)$ & 13  & 13 & 12\\
\hline
\end{tabular}

\vspace{12pt}

\begin{tabular}[t]{|l||c|c|c|c|}
\hline \parbox[l][20pt][c]{0pt}{}   $\alpha$ & 2 & $1+a$ &
$1+\frac{1}{a}$ &
\\ \hline
\parbox[l][20pt][c]{0pt}{} $\fc(\alpha)$ & 3 & 1 & 1 & 0 \\ \hline
\end{tabular}
\end{center}

\item\begin{tabular}{ll}$\g_{4,2}(1): \quad$ &$[e_1,e_4]=e_1,\
[e_2,e_4]=e_2, \ [e_3,e_4]=e_2+e_3 $
\end{tabular}
\vspace{-8pt}\begin{center}
\begin{tabular}[t]{|l||c|c|c|c|}
\hline \parbox[l][20pt][c]{0pt}{}   $\alpha$ & 1 &
\\ \hline
\parbox[l][20pt][c]{0pt}{} $\fa(\alpha)$ & 8 & 4  \\ \hline
\end{tabular}\qquad
\begin{tabular}[t]{|l||c|c|c|c|}
\hline \parbox[l][20pt][c]{0pt}{}   $\alpha$ & 2 &
\\ \hline
\parbox[l][20pt][c]{0pt}{} $\fb(\alpha)$ & 15 & 12  \\ \hline
\end{tabular}\qquad
\begin{tabular}[t]{|l||c|c|c|c|}
\hline \parbox[l][20pt][c]{0pt}{}   $\alpha$ & 2 &
\\ \hline
\parbox[l][20pt][c]{0pt}{} $\fc(\alpha)$ & 7 & 0  \\ \hline
\end{tabular}
\end{center}

\item\begin{tabular}{ll}$\g_{4,2}(-2): \quad$ &$[e_1,e_4]=-2e_1,\
[e_2,e_4]=e_2, \ [e_3,e_4]=e_2+e_3 $
\end{tabular}
\vspace{-8pt}\begin{center}
\begin{tabular}[t]{|l||c|c|c|c|}
\hline \parbox[l][20pt][c]{0pt}{}   $\alpha$ & 1 & $-2$&
$-\frac{1}{2}$& \\ \hline
\parbox[l][20pt][c]{0pt}{} $\fa(\alpha)$ & 6 & 5 & 5 & 4  \\ \hline
\end{tabular}\qquad
\begin{tabular}[t]{|l||c|c|c|c|}
\hline \parbox[l][20pt][c]{0pt}{}   $\alpha$ & -1 &
\\ \hline
\parbox[l][20pt][c]{0pt}{} $\fb(\alpha)$ & 15 & 12  \\ \hline
\end{tabular}\qquad
\begin{tabular}[t]{|l||c|c|c|c|}
\hline \parbox[l][20pt][c]{0pt}{}   $\alpha$ & 2 & $-1$ &
$\frac{1}{2}$ &
\\ \hline
\parbox[l][20pt][c]{0pt}{} $\fc(\alpha)$ & 3 & 1 & 1 & 0 \\ \hline
\end{tabular}
\end{center}

\item\begin{tabular}{ll}$\g_{4,2}(-1): \quad$ &$[e_1,e_4]=-e_1,\
[e_2,e_4]=e_2, \ [e_3,e_4]=e_2+e_3$
\end{tabular}
\vspace{-8pt}\begin{center}
\begin{tabular}[t]{|l||c|c|c|c|}
\hline \parbox[l][20pt][c]{0pt}{}   $\alpha$ & 1 & $-1$&
\\ \hline
\parbox[l][20pt][c]{0pt}{} $\fa(\alpha)$ & 6 & 6 &  4  \\ \hline
\end{tabular}\qquad
\begin{tabular}[t]{|l||c|c|c|}
\hline  \parbox[l][20pt][c]{0pt}{}  $\alpha$ & $-2$ & $0$ &
\\ \hline
\parbox[l][20pt][c]{0pt}{} $\fb(\alpha)$ & 13  & 16 & 12\\
\hline
\end{tabular}\qquad
\begin{tabular}[t]{|l||c|c|c|c|}
\hline \parbox[l][20pt][c]{0pt}{}   $\alpha$ & $0$ & $2$ &
\\ \hline
\parbox[l][20pt][c]{0pt}{} $\fc(\alpha)$ & 2 & 3 & 0  \\ \hline
\end{tabular}
\end{center}

\item\begin{tabular}{ll}$\g_{4,3}: \quad$ &$[e_1,e_4]=e_1,\
 [e_3,e_4]=e_2$
\end{tabular}
\vspace{-8pt}\begin{center}
\begin{tabular}[t]{|l||c|c|c|c|}
\hline \parbox[l][20pt][c]{0pt}{}   $\alpha$ & 1 & $0$&
\\ \hline
\parbox[l][20pt][c]{0pt}{} $\fa(\alpha)$ & 6 & 7 &  6  \\ \hline
\end{tabular}\qquad
\begin{tabular}[t]{|l||c|c|c|c|}
\hline \parbox[l][20pt][c]{0pt}{}   $\alpha$ & 0 &
\\ \hline
\parbox[l][20pt][c]{0pt}{} $\fb(\alpha)$ & 16 & 13  \\ \hline
\end{tabular}\qquad
\begin{tabular}[t]{|l||c|c|c|c|}
\hline \parbox[l][20pt][c]{0pt}{}   $\alpha$ & 1 & 2 &
\\ \hline
\parbox[l][20pt][c]{0pt}{} $\fc(\alpha)$ & 3 & 3& 2  \\ \hline
\end{tabular}
\end{center}

\item\begin{tabular}{ll}$\g_{4,4}: \quad$ &$[e_1,e_4]=e_1,\
[e_2,e_4]=e_1+e_2, \ [e_3,e_4]=e_2+e_3$
\end{tabular}
\vspace{-8pt}\begin{center}
\begin{tabular}[t]{|l||c|c|c|c|}
\hline \parbox[l][20pt][c]{0pt}{}   $\alpha$ & 1 &
\\ \hline
\parbox[l][20pt][c]{0pt}{} $\fa(\alpha)$ & 6 & 4 \\ \hline
\end{tabular}\qquad
\begin{tabular}[t]{|l||c|c|c|c|}
\hline \parbox[l][20pt][c]{0pt}{}   $\alpha$ & 2 &
\\ \hline
\parbox[l][20pt][c]{0pt}{} $\fb(\alpha)$ & 13 & 12  \\ \hline
\end{tabular}\qquad
\begin{tabular}[t]{|l||c|c|c|c|}
\hline \parbox[l][20pt][c]{0pt}{}   $\alpha$ & 1 & 2 &
\\ \hline
\parbox[l][20pt][c]{0pt}{} $\fc(\alpha)$ & 0 & 3& 0  \\ \hline
\end{tabular}
\end{center}

\item\label{g45ab}\begin{tabular}[t]{ll}$\g_{4,5}(a,b): \quad$ &$[e_1,e_4]=ae_1,\
[e_2,e_4]=be_2, \ [e_3,e_4]=e_3, $ \\ & $a\neq 0,\pm 1,\pm
b,1/b,b^2,-1-b, $   $\ b\neq 0,\pm 1,\pm a,1/a,a^2,-1-a$
\end{tabular}
\begin{center}
\begin{tabular}[t]{|l||c|c|c|c|c|c|c|c|}
\hline \parbox[l][20pt][c]{0pt}{}   $\alpha$ & 1 & $a$&
$\frac{1}{a}$& $b$& $\frac{1}{b}$ &  $\frac{a}{b}$ & $\frac{b}{a}$
&
\\ \hline
\parbox[l][20pt][c]{0pt}{} $\fa(\alpha)$ & 6 & 5 &  5& 5& 5&5&5&4  \\ \hline
\end{tabular}\qquad
\begin{tabular}[t]{|l||c|c|c|c|}
\hline \parbox[l][20pt][c]{0pt}{}   $\alpha$ &  $a+b$ &
$\frac{1+a}{b}$ & $\frac{1+b}{a}$ &
\\ \hline
\parbox[l][20pt][c]{0pt}{} $\fb(\alpha)$ & 13 & 13 & 13& 12  \\ \hline
\end{tabular}

\vspace{12pt}

\begin{tabular}[t]{|l||c|c|c|c|c|c|c|c|}
\hline \parbox[l][20pt][c]{0pt}{}   $\alpha$ & 2 & $1+a$& $1+b$ &
$1+\frac{1}{a}$&  $1+\frac{1}{b}$ & 1+ $\frac{a}{b}$ &
$1+\frac{b}{a}$ &
\\ \hline
\parbox[l][20pt][c]{0pt}{} $\fc(\alpha)$ & 3 & 1 &  1& 1& 1&1&1&0  \\ \hline
\end{tabular}
\end{center}

\item\label{g45am1ma}\begin{tabular}[t]{ll}$\g_{4,5}(a,-1-a): \quad$ &$[e_1,e_4]=ae_1,\
[e_2,e_4]=(-1-a)e_2, \ [e_3,e_4]=e_3, $ \\ & $a\neq 0,\pm 1,-2
,-1/2,-1/2\pm i\sqrt{3}/2  $
\end{tabular}
\begin{center}
\begin{tabular}[t]{|l||c|c|c|c|c|c|c|c|}
\hline \parbox[l][20pt][c]{0pt}{}   $\alpha$ & 1 & $a$&
$\frac{1}{a}$& $-1-a$& $-\frac{1}{1+a}$ &  $-\frac{a}{a+1}$ &
$-\frac{a+1}{a}$ &
\\ \hline
\parbox[l][20pt][c]{0pt}{} $\fa(\alpha)$ & 6 & 5 &  5& 5& 5&5&5&4  \\ \hline
\end{tabular}\qquad
\begin{tabular}[t]{|l||c|c|c|c|}
\hline \parbox[l][20pt][c]{0pt}{}   $\alpha$ &  -1 &
\\ \hline
\parbox[l][20pt][c]{0pt}{} $\fb(\alpha)$ & 15 &  12  \\ \hline
\end{tabular}

\vspace{12pt}

\begin{tabular}[t]{|l||c|c|c|c|c|c|c|c|}
\hline \parbox[l][20pt][c]{0pt}{}   $\alpha$ & 2 & $1+a$& $-a$ &
$1+\frac{1}{a}$&  $\frac{a}{a+1}$ & $\frac{1}{a+1}$ &
$-\frac{1}{a}$ &
\\ \hline
\parbox[l][20pt][c]{0pt}{} $\fc(\alpha)$ & 3 & 1 &  1& 1& 1&1&1&0  \\ \hline
\end{tabular}
\end{center}

\item\label{g45aas}\begin{tabular}[t]{ll}$\g_{4,5}(a,a^2): \quad$ &$[e_1,e_4]=ae_1,\
[e_2,e_4]=a^2e_2, \ [e_3,e_4]=e_3, $ \\ & $a\neq 0,\pm 1,\pm
i,-1/2\pm i\sqrt{3}/2 $
\end{tabular}
\begin{center}
\begin{tabular}[t]{|l||c|c|c|c|c|c|c|c|}
\hline \parbox[l][20pt][c]{0pt}{}   $\alpha$ & 1 & $a$&
$\frac{1}{a}$& $a^2$& $\frac{1}{a^2}$ &
\\ \hline
\parbox[l][20pt][c]{0pt}{} $\fa(\alpha)$ & 6 & 6 &  6& 5& 5&4  \\ \hline
\end{tabular}\qquad
\begin{tabular}[t]{|l||c|c|c|c|}
\hline \parbox[l][20pt][c]{0pt}{}   $\alpha$ &  $a+a^2$ &
$\frac{a+1}{a^2}$ & $\frac{a^2+1}{a}$ &
\\ \hline
\parbox[l][20pt][c]{0pt}{} $\fb(\alpha)$ & 13 & 13 & 13& 12  \\ \hline
\end{tabular}

\vspace{12pt}

\begin{tabular}[t]{|l||c|c|c|c|c|c|c|c|}
\hline \parbox[l][20pt][c]{0pt}{}   $\alpha$ & 2 & $1+a$& $1+a^2$
& $1+\frac{1}{a}$&  $1+\frac{1}{a^2}$ &
\\ \hline
\parbox[l][20pt][c]{0pt}{} $\fc(\alpha)$ & 3 & 2 &  1& 2& 1&0  \\ \hline
\end{tabular}
\end{center}

\item\label{g45a1}\begin{tabular}[t]{ll}$\g_{4,5}(a,1): \quad$ &$[e_1,e_4]=ae_1,\
[e_2,e_4]=e_2, \ [e_3,e_4]=e_3, $ \\ & $a\neq 0,\pm 1,-2$
\end{tabular}
\begin{center}
\begin{tabular}[t]{|l||c|c|c|c|}
\hline \parbox[l][20pt][c]{0pt}{}   $\alpha$ & 1 & $a$&
$\frac{1}{a}$& \\ \hline
\parbox[l][20pt][c]{0pt}{} $\fa(\alpha)$ & 8 & 6 & 6 & 4  \\ \hline
\end{tabular}\qquad
\begin{tabular}[t]{|l||c|c|c|}
\hline  \parbox[l][20pt][c]{0pt}{}  $\alpha$ & $1+a$ &
$\frac{2}{a}$ &
\\ \hline
\parbox[l][20pt][c]{0pt}{} $\fb(\alpha)$ & 15  & 13 & 12\\
\hline
\end{tabular}

\vspace{12pt}

\begin{tabular}[t]{|l||c|c|c|c|}
\hline \parbox[l][20pt][c]{0pt}{}   $\alpha$ & 2 & $1+a$ &
$1+\frac{1}{a}$ &
\\ \hline
\parbox[l][20pt][c]{0pt}{} $\fc(\alpha)$ & 7 & 2 & 2 & 0 \\ \hline
\end{tabular}
\end{center}

\item\label{g45am1}\begin{tabular}[t]{ll}$\g_{4,5}(a,-1): \quad$ &$[e_1,e_4]=ae_1,\
[e_2,e_4]=-e_2, \ [e_3,e_4]=e_3, $ \\ & $a\neq 0,\pm 1,\pm i$
\end{tabular}
\begin{center}
\begin{tabular}[t]{|l||c|c|c|c|c|c|c|c|}
\hline \parbox[l][20pt][c]{0pt}{}   $\alpha$ & 1 & $a$&
$\frac{1}{a}$& $-1$&   $-a$ & $-\frac{1}{a}$ &
\\ \hline
\parbox[l][20pt][c]{0pt}{} $\fa(\alpha)$ & 6 & 5 &  5& 6& 5&5&4  \\ \hline
\end{tabular}\qquad
\begin{tabular}[t]{|l||c|c|c|c|}
\hline \parbox[l][20pt][c]{0pt}{}   $\alpha$ &  $-1+a$ & $-1-a$ &
$0$ &
\\ \hline
\parbox[l][20pt][c]{0pt}{} $\fb(\alpha)$ & 13 & 13 & 16& 12  \\ \hline
\end{tabular}

\vspace{12pt}

\begin{tabular}[t]{|l||c|c|c|c|c|c|c|c|}
\hline \parbox[l][20pt][c]{0pt}{}   $\alpha$ & 2 & $1+a$& $0$ &
$1+\frac{1}{a}$ & $1-a$ & $1-\frac{1}{a}$ &
\\ \hline
\parbox[l][20pt][c]{0pt}{} $\fc(\alpha)$ & 3 & 1 &  2& 1&1&1&0  \\ \hline
\end{tabular}
\end{center}

\item\begin{tabular}[t]{ll}$\g_{4,5}(1,1): \quad$ &$[e_1,e_4]=e_1,\
[e_2,e_4]=e_2, \ [e_3,e_4]=e_3 $
\end{tabular}
\vspace{-8pt}\begin{center}
\begin{tabular}[t]{|l||c|c|c|c|}
\hline \parbox[l][20pt][c]{0pt}{}   $\alpha$ & 1 &
\\ \hline
\parbox[l][20pt][c]{0pt}{} $\fa(\alpha)$ & 12 & 4 \\ \hline
\end{tabular}\qquad
\begin{tabular}[t]{|l||c|c|c|c|}
\hline \parbox[l][20pt][c]{0pt}{}   $\alpha$ & 2 &
\\ \hline
\parbox[l][20pt][c]{0pt}{} $\fb(\alpha)$ & 18 & 12  \\ \hline
\end{tabular}\qquad
\begin{tabular}[t]{|l||c|c|c|c|}
\hline \parbox[l][20pt][c]{0pt}{}   $\alpha$ &  2 &
\\ \hline
\parbox[l][20pt][c]{0pt}{} $\fc(\alpha)$ & 18 &  0  \\ \hline
\end{tabular}
\end{center}

\item\begin{tabular}[t]{ll}$\g_{4,5}(-1,1): \quad$ &$[e_1,e_4]=-e_1,\
[e_2,e_4]=e_2, \ [e_3,e_4]=e_3 $
\end{tabular}
\vspace{-8pt}\begin{center}
\begin{tabular}[t]{|l||c|c|c|c|}
\hline \parbox[l][20pt][c]{0pt}{}   $\alpha$ & 1 & $-1$&
\\ \hline
\parbox[l][20pt][c]{0pt}{} $\fa(\alpha)$ & 8 & 8 & 4  \\ \hline
\end{tabular}\qquad
\begin{tabular}[t]{|l||c|c|c|}
\hline  \parbox[l][20pt][c]{0pt}{}  $\alpha$ & $0$ & $-2$ &
\\ \hline
\parbox[l][20pt][c]{0pt}{} $\fb(\alpha)$ & 20  & 13 & 12\\
\hline
\end{tabular}\qquad
\begin{tabular}[t]{|l||c|c|c|c|}
\hline \parbox[l][20pt][c]{0pt}{}   $\alpha$ & 2 & $0$ &
\\ \hline
\parbox[l][20pt][c]{0pt}{} $\fc(\alpha)$ & 7 & 4 & 0 \\ \hline
\end{tabular}
\end{center}

\item\begin{tabular}[t]{ll}$\g_{4,5}(-2,1): \quad$ &$[e_1,e_4]=-2e_1,\
[e_2,e_4]=e_2, \ [e_3,e_4]=e_3 $
\end{tabular}
\vspace{-8pt}\begin{center}
\begin{tabular}[t]{|l||c|c|c|c|}
\hline \parbox[l][20pt][c]{0pt}{}   $\alpha$ & 1 & $-2$&
$-\frac{1}{2}$& \\ \hline
\parbox[l][20pt][c]{0pt}{} $\fa(\alpha)$ & 8 & 6 & 6 & 4  \\ \hline
\end{tabular}\qquad
\begin{tabular}[t]{|l||c|c|c|}
\hline  \parbox[l][20pt][c]{0pt}{}  $\alpha$ & $-1$ &
\\ \hline
\parbox[l][20pt][c]{0pt}{} $\fb(\alpha)$ & 16   & 12\\
\hline
\end{tabular}\qquad
\begin{tabular}[t]{|l||c|c|c|c|}
\hline \parbox[l][20pt][c]{0pt}{}   $\alpha$ & 2 & $-1$ &
$\frac{1}{2}$ &
\\ \hline
\parbox[l][20pt][c]{0pt}{} $\fc(\alpha)$ & 7 & 2 & 2 & 0 \\ \hline
\end{tabular}
\end{center}

\item\begin{tabular}[t]{ll}$\g_{4,5}(-\frac{1}{2}+\frac{\sqrt{3}}{2}i,-\frac{1}{2}-\frac{\sqrt{3}}{2}i): \quad$ &$[e_1,e_4]=(-\frac{1}{2}+\frac{\sqrt{3}}{2}i)e_1,\
[e_2,e_4]=(-\frac{1}{2}-\frac{\sqrt{3}}{2})ie_2,$\\ &
$[e_3,e_4]=e_3
$
\end{tabular}
\begin{center}
\begin{tabular}[t]{|l||c|c|c|c|}
\hline \parbox[l][20pt][c]{0pt}{}   $\alpha$ & 1 &
$-\frac{1}{2}+\frac{\sqrt{3}}{2}i$&
$-\frac{1}{2}-\frac{\sqrt{3}}{2}i$& \\ \hline
\parbox[l][20pt][c]{0pt}{} $\fa(\alpha)$ & 6 & 7 & 7 & 4  \\ \hline
\end{tabular}\qquad
\begin{tabular}[t]{|l||c|c|c|}
\hline  \parbox[l][20pt][c]{0pt}{}  $\alpha$ & $-1$ &
\\ \hline
\parbox[l][20pt][c]{0pt}{} $\fb(\alpha)$ & 15   & 12\\
\hline
\end{tabular}\qquad

\vspace{12pt}

\begin{tabular}[t]{|l||c|c|c|c|}
\hline \parbox[l][20pt][c]{0pt}{}   $\alpha$ & 2 &
$\frac{1}{2}-\frac{\sqrt{3}}{2}i$ &
$\frac{1}{2}+\frac{\sqrt{3}}{2}i$ &
\\ \hline
\parbox[l][20pt][c]{0pt}{} $\fc(\alpha)$ & 3 & 3 & 3 & 0 \\ \hline
\end{tabular}
\end{center}

\item\begin{tabular}[t]{ll}$\g_{4,5}(i,-1): \quad$ &$[e_1,e_4]=ie_1,\
[e_2,e_4]=-e_2, \ [e_3,e_4]=e_3 $
\end{tabular}
\vspace{-8pt}\begin{center}
\begin{tabular}[t]{|l||c|c|c|c|c|c|c|c|}
\hline \parbox[l][20pt][c]{0pt}{}   $\alpha$ & 1 & $i$& $-i$&
$-1$&
\\ \hline
\parbox[l][20pt][c]{0pt}{} $\fa(\alpha)$ & 6 & 6 &  6& 6&4  \\ \hline
\end{tabular}\qquad
\begin{tabular}[t]{|l||c|c|c|c|}
\hline \parbox[l][20pt][c]{0pt}{}   $\alpha$ &  $-1+i$ & $-1-i$ &
$0$ &
\\ \hline
\parbox[l][20pt][c]{0pt}{} $\fb(\alpha)$ & 13 & 13 & 16& 12  \\ \hline
\end{tabular}

\vspace{12pt}

\begin{tabular}[t]{|l||c|c|c|c|c|c|c|c|}
\hline \parbox[l][20pt][c]{0pt}{}   $\alpha$ & 2 & $1+i$& $0$ &
$1-i$  &
\\ \hline
\parbox[l][20pt][c]{0pt}{} $\fc(\alpha)$ & 3 & 2 &  2& 2&0  \\ \hline
\end{tabular}
\end{center}

\item\begin{tabular}[t]{ll}$\g_{4,7}: \quad$ &$[e_2,e_3]=e_1,\
[e_1,e_4]=2e_1, \ [e_2,e_4]=e_2, \ [e_3,e_4]=e_2+e_3$
\end{tabular}
\vspace{-8pt}\begin{center}
\begin{tabular}[t]{|l||c|c|c|c|}
\hline \parbox[l][20pt][c]{0pt}{}   $\alpha$ & 1 & $2$&
\\ \hline
\parbox[l][20pt][c]{0pt}{} $\fa(\alpha)$ & 5 & 4 & 3  \\ \hline
\end{tabular}\qquad
\begin{tabular}[t]{|l||c|c|c|c|c|c|c|c|}
\hline \parbox[l][20pt][c]{0pt}{}   $\alpha$ & 0 & $1$& $3$  &
\\ \hline
\parbox[l][20pt][c]{0pt}{} $\fc(\alpha)$ & 12 & 12 &  12& 11  \\ \hline
\end{tabular}\qquad
\begin{tabular}[t]{|l||c|c|c|c|}
\hline \parbox[l][20pt][c]{0pt}{}   $\alpha$ & $\frac{3}{2}$
 & $2$&
\\ \hline
\parbox[l][20pt][c]{0pt}{} $\fc(\alpha)$ & 1 & 1 & 0 \\ \hline
\end{tabular}
\end{center}

\item\label{g48}\begin{tabular}[t]{ll}$\g_{4,8}(a): \quad$ &$[e_2,e_3]=e_1,\
[e_1,e_4]=(1+a)e_1, \ [e_2,e_4]=e_2, \ [e_3,e_4]=ae_3$ \\ & $a\neq
0,\pm 1,\pm 2,\pm 1/2,-1/2\pm\sqrt{3}i/2 $
\end{tabular}
\begin{center}
\begin{tabular}[t]{|l||c|c|c|c|c|c|c|c|}
\hline \parbox[l][20pt][c]{0pt}{}   $\alpha$ & 1 & $2$&   $a$&
$\frac{1}{a}$&
\\ \hline
\parbox[l][20pt][c]{0pt}{} $\fa(\alpha)$ & 5 & 4 &  4& 4& 3  \\ \hline
\end{tabular}\qquad
\begin{tabular}[t]{|l||c|c|c|c|c|c|c|c|}
\hline \parbox[l][20pt][c]{0pt}{}   $\alpha$ & 0 & $1$& $1+2a$ &
$1+\frac{2}{a}$ & $\frac{5}{3}+\frac{2}{3}(a+\frac{1}{a})$ &
\\ \hline
\parbox[l][20pt][c]{0pt}{} $\fb(\alpha)$ & 12 & 12 &  12& 12&12&11  \\ \hline
\end{tabular}

\vspace{12pt}

\begin{tabular}[t]{|l||c|c|c|c|c|c|c|c|}
\hline \parbox[l][20pt][c]{0pt}{}   $\alpha$ & 2 &
$\frac{1+2a}{1+a}$& $\frac{2+a}{1+a}$&
\\ \hline
\parbox[l][20pt][c]{0pt}{} $\fc(\alpha)$ & 1 & 1 &  1&0  \\ \hline
\end{tabular}
\end{center}

\item\begin{tabular}[t]{ll}$\g_{4,8}(1): \quad$ &$[e_2,e_3]=e_1,\
[e_1,e_4]=2e_1, \ [e_2,e_4]=e_2, \ [e_3,e_4]=e_3$
\end{tabular}
\vspace{-8pt}\begin{center}
\begin{tabular}[t]{|l||c|c|c|c|c|c|c|c|}
\hline \parbox[l][20pt][c]{0pt}{}   $\alpha$ & 1 & $2$&
\\ \hline
\parbox[l][20pt][c]{0pt}{} $\fa(\alpha)$ & 7 & 4 &  3  \\ \hline
\end{tabular}\qquad
\begin{tabular}[t]{|l||c|c|c|c|c|c|c|c|}
\hline \parbox[l][20pt][c]{0pt}{}   $\alpha$ & 0 & $1$& $3$ &
\\ \hline
\parbox[l][20pt][c]{0pt}{} $\fb(\alpha)$ & 12 & 12 &  14&11  \\ \hline
\end{tabular}\qquad
\begin{tabular}[t]{|l||c|c|c|c|c|c|c|c|}
\hline \parbox[l][20pt][c]{0pt}{}   $\alpha$ & 2 & $\frac{3}{2}$&
\\ \hline
\parbox[l][20pt][c]{0pt}{} $\fc(\alpha)$ & 1 & 2 & 0  \\ \hline
\end{tabular}
\end{center}

\item\begin{tabular}[t]{ll}$\g_{4,8}(2): \quad$ &$[e_2,e_3]=e_1,\
[e_1,e_4]=3e_1, \ [e_2,e_4]=e_2, \ [e_3,e_4]=2e_3$
\end{tabular}
\vspace{-8pt}\begin{center}
\begin{tabular}[t]{|l||c|c|c|c|c|c|c|c|}
\hline \parbox[l][20pt][c]{0pt}{}   $\alpha$ & 1 & $2$&
$\frac{1}{2}$&
\\ \hline
\parbox[l][20pt][c]{0pt}{} $\fa(\alpha)$ & 5 & 5 &  4&  3  \\ \hline
\end{tabular}\qquad
\begin{tabular}[t]{|l||c|c|c|c|c|c|c|c|}
\hline \parbox[l][20pt][c]{0pt}{}   $\alpha$ & 0 & $1$& $5$ & $2$
& $\frac{10}{3}$ &
\\ \hline
\parbox[l][20pt][c]{0pt}{} $\fb(\alpha)$ & 12 & 12 &  12& 12&12&11  \\ \hline
\end{tabular}
\vspace{12pt}

\begin{tabular}[t]{|l||c|c|c|c|c|c|c|c|}
\hline \parbox[l][20pt][c]{0pt}{}   $\alpha$ & 2 & $\frac{5}{3}$&
$\frac{4}{3}$&
\\ \hline
\parbox[l][20pt][c]{0pt}{} $\fc(\alpha)$ & 1 & 1 &  1&0  \\ \hline
\end{tabular}
\end{center}

\item\begin{tabular}[t]{ll}$\g_{4,8}(0): \quad$ &$[e_2,e_3]=e_1,\
[e_1,e_4]=e_1, \ [e_2,e_4]=e_2$

\end{tabular}
\vspace{-8pt}\begin{center}
\begin{tabular}[t]{|l||c|c|c|c|c|c|c|c|}
\hline \parbox[l][20pt][c]{0pt}{}   $\alpha$ & 1 & $0$&
\\ \hline
\parbox[l][20pt][c]{0pt}{} $\fa(\alpha)$ & 5 & 6 & 4  \\ \hline
\end{tabular}\qquad
\begin{tabular}[t]{|l||c|c|c|c|c|c|c|c|}
\hline \parbox[l][20pt][c]{0pt}{}   $\alpha$ & 0 & $1$&
\\ \hline
\parbox[l][20pt][c]{0pt}{} $\fb(\alpha)$ & 12 & 13 & 11  \\ \hline
\end{tabular}\qquad
\begin{tabular}[t]{|l||c|c|c|c|c|c|c|c|}
\hline \parbox[l][20pt][c]{0pt}{}   $\alpha$ & 1 & 2&
\\ \hline
\parbox[l][20pt][c]{0pt}{} $\fc(\alpha)$ & 2 & 2 &0  \\ \hline
\end{tabular}
\end{center}

\item\begin{tabular}[t]{ll}$\g_{4,8}(-1): \quad$ &$[e_2,e_3]=e_1,\
\ [e_2,e_4]=e_2, \ [e_3,e_4]=-e_3$
\end{tabular}
\vspace{-8pt}\begin{center}
\begin{tabular}[t]{|l||c|c|c|c|c|c|c|c|}
\hline \parbox[l][20pt][c]{0pt}{}   $\alpha$ & 1 & $-1$&
\\ \hline
\parbox[l][20pt][c]{0pt}{} $\fa(\alpha)$ & 5 & 6 & 4  \\ \hline
\end{tabular}\qquad
\begin{tabular}[t]{|l||c|c|c|c|c|c|c|c|}
\hline \parbox[l][20pt][c]{0pt}{}   $\alpha$ & 1 & $-1$&
\\ \hline
\parbox[l][20pt][c]{0pt}{} $\fb(\alpha)$ & 13 & 14 & 12  \\ \hline
\end{tabular}\qquad
\begin{tabular}[t]{|l||c|c|c|c|c|c|c|c|}
\hline \parbox[l][20pt][c]{0pt}{}   $\alpha$ & 2 &
\\ \hline
\parbox[l][20pt][c]{0pt}{} $\fc(\alpha)$ & 1 & 0  \\ \hline
\end{tabular}
\end{center}

\item\begin{tabular}[t]{ll}$\g_{4,8}(-2): \quad$ &$[e_2,e_3]=e_1,\
[e_1,e_4]=-e_1, \ [e_2,e_4]=e_2, \ [e_3,e_4]=-2e_3$
\end{tabular}
\vspace{-8pt}\begin{center}
\begin{tabular}[t]{|l||c|c|c|c|c|c|c|c|}
\hline \parbox[l][20pt][c]{0pt}{}   $\alpha$ & 1 & $2$&   $-2$&
$-\frac{1}{2}$&
\\ \hline
\parbox[l][20pt][c]{0pt}{} $\fa(\alpha)$ & 5 & 4 &  4& 4& 3  \\ \hline
\end{tabular}\qquad
\begin{tabular}[t]{|l||c|c|c|c|c|c|c|c|}
\hline \parbox[l][20pt][c]{0pt}{}   $\alpha$ & 0 & $1$& $-3$ &
\\ \hline
\parbox[l][20pt][c]{0pt}{} $\fb(\alpha)$ & 16 & 12 &  12& 11  \\ \hline
\end{tabular}
\vspace{12pt}

\begin{tabular}[t]{|l||c|c|c|c|c|c|c|c|}
\hline \parbox[l][20pt][c]{0pt}{}   $\alpha$ & 2 & $3$& $0$&
\\ \hline
\parbox[l][20pt][c]{0pt}{} $\fc(\alpha)$ & 1 & 1 &  1&0  \\ \hline
\end{tabular}
\end{center}

\item\begin{tabular}[t]{ll}$\g_{4,8}(-\frac{1}{2}+\frac{\sqrt{3}}{2}i): \quad$ &$[e_2,e_3]=e_1,\
[e_1,e_4]=(\frac{1}{2}+\frac{\sqrt{3}}{2}i)e_1, \ [e_2,e_4]=e_2$,
\\ & $[e_3,e_4]=(-\frac{1}{2}+\frac{\sqrt{3}}{2}i)e_3$
\end{tabular}
\vspace{-0pt}\begin{center}
\begin{tabular}[t]{|l||c|c|c|c|c|c|c|c|}
\hline \parbox[l][20pt][c]{0pt}{}   $\alpha$ & 1 & $2$&
$-\frac{1}{2}+\frac{\sqrt{3}}{2}i$&
$-\frac{1}{2}-\frac{\sqrt{3}}{2}i$&
\\ \hline
\parbox[l][20pt][c]{0pt}{} $\fa(\alpha)$ & 5 & 4 &  4& 4& 3  \\ \hline
\end{tabular}\qquad
\begin{tabular}[t]{|l||c|c|c|c|c|c|c|c|}
\hline \parbox[l][20pt][c]{0pt}{}   $\alpha$ & 0 & $1$&
$\sqrt{3}i$ & $-\sqrt{3}i$ &
\\ \hline
\parbox[l][20pt][c]{0pt}{} $\fb(\alpha)$ & 12 &  12& 12&12&11  \\ \hline
\end{tabular}
\vspace{12pt}

\begin{tabular}[t]{|l||c|c|c|c|c|c|c|c|}
\hline \parbox[l][20pt][c]{0pt}{}   $\alpha$ & 2 &
$\frac{3}{2}+\frac{\sqrt{3}}{2}i$&
$\frac{3}{2}-\frac{\sqrt{3}}{2}i$&
\\ \hline
\parbox[l][20pt][c]{0pt}{} $\fc(\alpha)$ & 1 & 1 &  1&0  \\ \hline
\end{tabular}
\end{center}
\end{enumerate}
                           }
  \section*{Appendix B: Proof of Theorem \ref{class4dim} }
\begin{lemma}\label{lemg441}
For the following complex four--dimensional Lie algebras defined in
Appendix A it holds:
\begin{itemize}\item[]
\begin{itemize}
\item[(g-\ref{g34})] $\g_{3,4}(a)\oplus\g_1,\, a\neq 0,\pm 1$
\begin{align*} \fa\g_{3,4}(a)\oplus\g_1:\, & 7_1,6_3,5\\
\fb\g_{3,4}(a)\oplus\g_1:\, & 13_3,12
\end{align*}
\item[(g-\ref{g42})] $\g_{4,2}(a),\, a\neq 0,\pm 1,-2$
\begin{align*}\fa\g_{4,2}(a):\,& 6_1,5_2,4\\
\fb\g_{4,2}(a):\,&13_2,12
 \end{align*}
\item[(g-\ref{g45ab})] $\g_{4,5}(a,b),\, a\neq 0,\pm 1,\pm
b,\frac{1}{b},b^2,-1-b, $   $\ b\neq 0,\pm 1,\pm
a,\frac{1}{a},a^2,-1-a$
\begin{align*}\fa\g_{4,5}(a,b):\,& 6_1,5_6,4\\
\fb\g_{4,5}(a,b):\,&13_3,12
 \end{align*}
\item[(g-\ref{g45am1ma})] $\g_{4,5}(a,-1-a),\, a\neq 0,\pm 1,-2,-\frac{1}{2},-\frac{1}{2}\pm \frac{\sqrt{3}}{2}i$
\begin{align*} \fa\g_{4,5}(a,-1-a):\, & 6_1,5_6,4\\
\fb\g_{4,5}(a,-1-a):\, & 15_1,12
\end{align*}
\item[(g-\ref{g45aas})] $\g_{4,5}(a,a^2),\, a\neq 0,\pm 1,\pm i,-\frac{1}{2}\pm \frac{\sqrt{3}}{2}i$
\begin{align*} \fa\g_{4,5}(a,a^2):\, & 6_3,5_2,4\\
\fb\g_{4,5}(a,a^2):\, & 13_3,12
\end{align*}
\item[(g-\ref{g45a1})] $\g_{4,5}(a,1),\, a\neq 0,\pm 1,-2$
\begin{align*} \fa\g_{4,5}(a,1):\, & 8_1,6_2,4\\
\fb\g_{4,5}(a,1):\, & 15_1,13_1,12
\end{align*}
\item[(g-\ref{g45am1})] $\g_{4,5}(a,-1),\, a\neq 0,\pm 1,\pm i$
\begin{align*} \fa\g_{4,5}(a,-1):\, & 6_2,5_4,4\\
\fb\g_{4,5}(a,-1):\, & 16_1,13_2,12
\end{align*}
\item[(g-\ref{g48})] $\g_{4,8}(a),\, a\neq 0,\pm 1,\pm 2,\pm \frac{1}{2},-\frac{1}{2}\pm \frac{\sqrt{3}}{2}i$
\begin{align*} \fa\g_{4,8}(a):\, & 5_1,4_3,3\\
\fb\g_{4,8}(a):\, & 12_5,11
\end{align*}
\end{itemize}
\end{itemize}
\end{lemma}
\begin{proof}
Let us give the detailed proof for the case (g-18). We have to check for solutions each of 15 possible equalities
\begin{align*}
a=&\frac{1}{a}\\ a=&-1-a\\ a=&-\frac{a}{a+1}\\ &\vdots \\
-\frac{a}{1+a}=&-\frac{a+1}{a}.
\end{align*}
These equations have all solutions in the set $\left\{0,\pm 1,-2
,-\frac{1}{2},-\frac{1}{2}\pm i\frac{\sqrt{3}}{2}\right\}$ --
these values we excluded from the beginning.
The rest of the proof is analogous.
\end{proof}

     \begin{lemma}\label{lemg44}
For the four--dimensional Lie algebras from Lemma \ref{lemg441} it
holds:
\begin{itemize}\item[]
\begin{itemize}
\item[(g-\ref{g34})] If $\fa\g_{3,4}(a)\oplus \g_1=\fa\g_{3,4}(a')\oplus \g_1$ then $a'=a,\frac{1}{a}$.
\item[(g-\ref{g42})] If $\fb\g_{4,2}(a)=\fb\g_{4,2}(a')$ then $a'=a$.
\item[(g-\ref{g45ab})] If $\fb\g_{4,5}(a,b)=\fb\g_{4,5}(a',b')$ then $$(a',b')=(a,b),(b,a),\left(\frac{1}{a},\frac{b}{a}\right),\left(\frac{b}{a},\frac{1}{a}\right),
\left(\frac{1}{b},\frac{a}{b}\right),\left(\frac{a}{b},\frac{1}{b}\right)
.$$
\item[(g-\ref{g45am1ma})] If $\fa\g_{4,5}(a,-1-a)=\fa\g_{4,5}(a',-1-a')$ then
$$a'=a,\frac{1}{a},-\frac{a}{1+a},-1-\frac{1}{a},-1-a,-\frac{1}{1+a}.$$
\item[(g-\ref{g45aas})] If $\fa\g_{4,5}(a,a^2)=\fa\g_{4,5}(a',{a'}^2) $ then
$a'=a,\frac{1}{a}$.
\item[(g-\ref{g45a1})] If $\fb\g_{4,5}(a,1)=\fb\g_{4,5}(a',1)$ then $a'=a$.
\item[(g-\ref{g45am1})] If $\fb\g_{4,5}(a,-1)=\fb\g_{4,5}(a',-1)$ then $a'=a,-a$.
\item[(g-\ref{g48})] If $\fa\g_{4,8}(a)=\fa\g_{4,8}(a')$ then $a'=a,\frac{1}{a}$.
\end{itemize}
\end{itemize}
\end{lemma}
\begin{proof}
{\it Cases} (g-\ref{g34}), (g-\ref{g45aas}), (g-\ref{g45a1}),
(g-\ref{g45am1})$\,$ and$\,$ (g-\ref{g48}) are obvious.

{\it Case} (g-\ref{g42}). The function $\fb$ of $\g_{4,2}(a')$ has
the form
\begin{equation}
\begin{tabular}[c]{|l||c|c|c|}
\hline  \parbox[l][20pt][c]{0pt}{}  $\alpha$ & $1+a'$ &
$\frac{2}{a'}$ &
\\ \hline
\parbox[l][20pt][c]{0pt}{} $\fb\g_{4,2}(a')(\alpha)$ & 13  & 13 & 12\\
\hline
\end{tabular}
\end{equation}
and there are two possibilities:
\begin{itemize}\item[]
\begin{itemize}
\item[(12)] If $a+1=a'+1, \, \frac{2}{a}=\frac{2}{a'}$ then
$a'=a$.
\item[(21)] If $a+1=\frac{2}{a'},\, a'+1=\frac{2}{a}$ then
$a=a'=1,-2$, which is not possible.
\end{itemize}
\end{itemize}

  {\it Case} (g-\ref{g45ab}). The function $\fb$ of
$\g_{4,5}(a',b')$ has the form
\begin{equation}\label{tablegen2}
\begin{tabular}[c]{|l||c|c|c|c|}
\hline \parbox[l][20pt][c]{0pt}{}   $\alpha$ &  $a'+b'$ &
$\frac{1+a'}{b'}$ & $\frac{1+b'}{a'}$ &
\\ \hline
\parbox[l][20pt][c]{0pt}{} $\fb\g_{4,5}(a',b')(\alpha)$ & 13 & 13 & 13& 12  \\ \hline
\end{tabular}
\end{equation}
and there are six possible correspondences between this table and
(g-17). We obtain:\begin{itemize}\item[]\begin{itemize}
\item[(123)] If $z_2=\frac{1+b}{a},\, z_3=a+b$ then $a'=a,b'=b$.
\item[(213)] If $z_2=\frac{1+a}{b},\, z_3=a+b$ then $a'=b,b'=a$.
\item[(132)] If $z_2=a+b,\, z_3=\frac{1+b}{a}$ then $a'=\frac{1}{a},b'=\frac{b}{a}$.
\item[(312)] If $z_2=\frac{1+a}{b},\, z_3=\frac{1+b}{a}$ then $a'=\frac{b}{a},b'=\frac{1}{a}$.
\item[(231)] If $z_2=a+1,\, z_3=\frac{1+a}{b}$ then $a'=\frac{1}{b},b'=\frac{a}{b}$.
\item[(312)] If $z_2=\frac{1+b}{a},\, z_3=\frac{1+a}{b}$ then $a'=\frac{a}{b},b'=\frac{1}{b}$.
\end{itemize}
\end{itemize}

{\it Case} (g-18). It is convenient to note that six values in the
table (g-18) can be arranged in the triple of pairs
$\{a,\frac{1}{a}\}$, $\{-1-a, -\frac{1}{1+a} \}$ and
$\{-\frac{1+a}{a}, -\frac{a}{1+a} \}$. Then one checks directly
only $6\cdot 2^3=48$ permutations and obtains the solutions like
in the previous case.
\end{proof}

\begin{cor}\label{corr44} For the four--dimensional Lie algebras from Lemma \ref{lemg441} it
holds:\begin{itemize}\item[]
\begin{itemize}
\item[(g-\ref{g34})] If $\fa\g_{3,4}(a)\oplus \g_1=\fa\g_{3,4}(a')\oplus \g_1$ then $\g_{3,4}(a)\oplus\g_1\cong\g_{3,4}(a')\oplus \g_1$.
\item[(g-\ref{g45ab})] If $\fb\g_{4,5}(a,b)=\fb\g_{4,5}(a',b')$ then
$\g_{4,5}(a,b)\cong\g_{4,5}(a',b')$.
\item[(g-\ref{g45am1ma})] If $\fa\g_{4,5}(a,-1-a)=\fa\g_{4,5}(a',-1-a')$ then
$\g_{4,5}(a,-1-a)\cong\g_{4,5}(a',-1-a')$.
\item[(g-\ref{g45aas})] If $\fa\g_{4,5}(a,a^2)=\fa\g_{4,5}(a',{a'}^2) $ then
$\g_{4,5}(a,a^2)\cong\g_{4,5}(a',{a'}^2) $.
\item[(g-\ref{g45am1})] If $\fb\g_{4,5}(a,-1)=\fb\g_{4,5}(a',-1)$ then $\g_{4,5}(a,-1)\cong\g_{4,5}(a',-1)$.
\item[(g-\ref{g48})] If $\fa\g_{4,8}(a)=\fa\g_{4,8}(a')$ then $\g_{4,8}(a)\cong\g_{4,8}(a')$.
\end{itemize}
\end{itemize}
\end{cor}
\begin{proof}
The statement follows from Lemma \ref{lemg44} and from the
relations
\begin{align}
\g_{3,4}(a)\oplus\g_1& \cong  \g_{3,4}(1/a)\oplus \g_1 \\
\g_{4,5}(a,b)\cong\g_{4,5}(b,a)
\cong\g_{4,5}\left(\frac{1}{a},\frac{b}{a}\right) &
\cong\g_{4,5}\left(\frac{b}{a},\frac{1}{a}\right)\cong\g_{4,5}
\left(\frac{1}{b},\frac{a}{b}\right)\cong\g_{4,5}\left(\frac{a}{b},\frac{1}{b}\right)\\
\g_{4,8}(a)&\cong\g_{4,8}(1/a),
\end{align}
which hold for all $a,b\neq 0$ and can be directly verified.
\end{proof}
{\it Proof of Theorem \ref{class4dim}: }

$\Rightarrow:$ See Corollary~\ref{invarfunc2}.

$\Leftarrow:$ According to Lemmas~\ref{lemg441}, \ref{lemg44},
Corollary~\ref{corr44} and observing the tables in
Appendix~A, we conclude that all non--isomorphic
four-dimensional complex Lie algebras differ at least in one of
the functions $\fa$ or $\fb$.


\end{document}